\documentclass[11pt]{article}

\usepackage[table]{xcolor}
\usepackage{graphicx,pslatex,float,color,array,amssymb,amsmath,euscript}
\usepackage[paperwidth=21.0cm,paperheight=29.7cm,textwidth=16cm,textheight=24.5cm,top=3cm,left=2.5cm]{geometry}
\usepackage[utf8]{inputenc}
\usepackage[OT1]{fontenc}
\definecolor{navy}{rgb}{0,0,0.5}
\usepackage[%
           pdftex,
           hyperindex=true,
           colorlinks=true,
           linkcolor=navy,
           anchorcolor=magenta,
           citecolor=navy,
           urlcolor=navy,
           unicode,
           implicit=true]{hyperref}

  \definecolor{navy}{rgb}{0,0,0.5}
  
  \renewcommand{\vec}[1]{\boldsymbol{#1}}

  \newcommand{\be}{\begin{equation}}
  \newcommand{\ee}{\end{equation}}
  
  \usepackage{url,bm,lineno,xspace,soul,ulem,wasysym}
  \definecolor{mygreen}{rgb}{0,0.5,0.}
  \definecolor{myblue}{rgb}{0,0.3,0.8}
  \definecolor{myred}{rgb}{0.65,0.2,0}
  \definecolor{orange}{rgb}{0.8,0.4,0.0}

\usepackage[table]{xcolor}
\usepackage{pgf}
\usepackage{fancyhdr}
\usepackage{lastpage}

\let\emph\textit

\title{Electrical and Thermal Conductivity of\\ Complex-Shaped Contact Spots}

\author{\large Paul Beguin \& Vladislav A. Yastrebov\href{https://orcid.org/0000-0002-4052-3557}{\includegraphics[height=10pt]{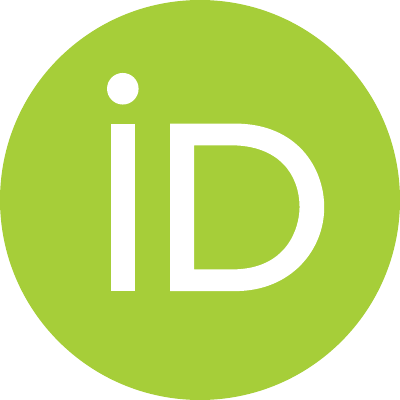}}}
\date{\footnotesize\textit{MINES Paris - PSL, Centre des Mat\'eriaux, CNRS UMR 7633, BP 87, 91003, Evry, France}}

\begin{document}

\maketitle

\begin{abstract}
  \noindent   This paper explores the electrical and thermal conductivity of complex contact spots on the surface of a half-space. 
  Employing an in-house Fast Boundary Element Method implementation, various complex geometries were studied. 
  Our investigation begins with annulus contact spots to assess the impact of connectedness. 
  We then study shape effects on "multi-petal" spots exhibiting dihedral symmetry, resembling flowers, stars, and gears. 
  The analysis culminates with self-affine shapes, representing a multiscale generalization of the multi-petal forms.
  In each case, we introduce appropriate normalizations and develop phenomenological models. 
  For multi-petal shapes, our model relies on a single geometric parameter: the normalized number of "petals".
  This approach inspired the form of the phenomenological model for self-affine spots,
  which maintains physical consistency and relies on four geometric characteristics: standard deviation, second spectral moment, Nayak parameter, and Hurst exponent. 
  As a by product, these models enabled us to suggest flux estimations for an infinite number of petals and the fractal limit. 
  This study represents an initial step into understanding the conductivity of complex contact interfaces, which commonly occur in the contact of rough surfaces.

\begin{flushleft}
\end{flushleft}

\noindent Supplementary material for this article is supplied as a separate archive available from Zenodo \href{https://doi.org/10.5281/zenodo.10200997}{10.5281/zenodo.10200997}. The fast BEM code used for simulations is shared as open-source at \href{https://github.com/vyastreb/HBEM}{github.com/vyastreb/HBEM} and a snapshot used for all simulations in the paper is available at Software Heritage \href{https://archive.softwareheritage.org/swh:1:dir:7edf21c11efe9437705cc2d9edfc096f8a0201b4;origin=https://github.com/vyastreb/HBEM;visit=swh:1:snp:d5ab4d2e357dd87c10738f1460e7b85e6997b4ca;anchor=swh:1:rev:1e9d38bc28780becd714a2ab015765c0b1266cd8}{SWHID}.

\vspace{10pt}
\hrule
\end{abstract}

\tableofcontents

\section{Introduction}

The study of mechanical contact plays a crucial role in numerous natural and engineering systems.
Some of these systems involve sliding motion, frictional resistance, lubrication and wear, while others operate exclusively in normal contact.
In the former case, the interplay of mechanical deformation, chemistry, and thermal effects due to frictional heat generation and diffusion significantly influence the overall behavior of the interface~\cite{blok1963flash,bowden2001friction,rice2006heating}.
This heat generation, coupled with intense shear deformation, can lead to phase transitions/transformations, recrystallization, and other metallurgical or chemical effects~\cite{jacobson2009surface,goldsby2011flash,yamashita2015scale}, such as local welding/galling and abrasive or adhesive wear, depending on oxygen influx~\cite{baydoun2020modelling,baydoun2022explicit}.
In the case of normal contact, frictional dissipation within the interface remains minimal.
However, the nature of contact interfaces profoundly impacts their conductive properties for both heat and electric charge.

In both natural and engineering systems, all surfaces exhibit roughness.
As a result, at relatively light contact loads, the true contact area -- formed by several intimately contacting "asperities" -- is smaller than the apparent contact area between solids.
Apart from other mechanical and physical properties, the contact area fraction and its morphology also dictate energy transfer through the contact interface.
Since the contact area fraction evolves under increasing load~\cite{archard1953contact,archard1957elastic,greenwood1966contact}, the interface conductivity depends on this load as well.

Surface roughness can be described by its height distribution or its moments.
In the simplest case, surfaces can be described as Gaussian, but in practice, most surfaces in contact exhibit an asymmetric height distribution due to wear or residual plastic deformations induced during contact, particularly after the running-in process.
An example of this asymmetry is asphalt concrete, which, due to manufacturing techniques, inherently has relatively flat plateaus and deep valleys.
Although height distribution or first moments of roughness are relevant parameters for specific applications, they do not adequately describe mechanical contact between rough surfaces, where the curvature of contacting "asperities" is of primary importance.
Consequently, power spectral density (PSD) and its first moments represent the key parameters in describing the elastic contact of rough surfaces~\cite{greenwood1966contact,nayak1971random,persson2001theory,yastrebov2017role}.

Aligned with PSD-based models which provide a multiscale representation of surface features, some authors proposed \textit{fractal models} of surface roughness~\cite{majumdar1990fractal,majumdar1991fractal,dodds1973description}, which, to some extent also ensure a suitable description of real rough surfaces in the framework of Archard's concept of "protuberances on protuberances on protuberances"~\cite{archard1953contact} however the mechanical consistency of such models dealing with elasto-plastic contact remains questionable.

In the study of rough surfaces in contact, one can adopt either statistical or deterministic approaches.
The former, which includes multi-asperity models~\cite{greenwood1966contact,bush1975elastic,greenwood2006simplified,carbone2008asperity} and Persson's model~\cite{persson2001theory,manners2006some}, deals with probability densities and average statistical properties of roughness.
This approach is applicable to contacts with a statistically meaningful number of contacting asperities.

These analytical and semi-analytical methods are built on certain assumptions and, as such, have limitations when it comes to predicting the rigorously formulated contact problem between rough surfaces~\cite{manners2006some,carbone2008asperity,yastrebov2017role}.
On the other hand, deterministic approaches, such as multi-asperity models with interactions~\cite[e.g.]{afferrante2012interacting,yastrebov2019elastic} can effectively account for a small number of contact spots, their interaction and spatial heterogeneities in roughness.
Different more advanced methods can be employed to address contact problems within a deterministic framework.
The most versatile is the finite-element method~\cite{hyun2004finite,pei2005finite,yastrebov2011rough,gao2006elastic}, which can inherently handle nonlinear and heterogeneous material behavior and large deformations.
However, it is computationally intensive and necessitates solving mechanical equations not only on the surface but also in the bulk.
Another group of methods, including boundary-element and spectral methods, focuses exclusively on surface interactions and, in their basic implementation, relies on space-invariant fundamental solutions~\cite{kalker1977variational,stanley1997fft,polonsky1999numerical,bonnet1999boundary,campana2006practical}.
Over the past few decades, these methods have been extended to tackle more complex problems involving heterogeneous and nonlinear materials~\cite{putignano2015mechanics,amuzuga2016fully,frerot2019fourier,perez2023interplay}.

When two conductive solids come into contact, a localized resistance emerges due to the discontinuous nature of the actual contact area.
From a mathematical standpoint, at the macro-scale level, this phenomenon is characterized by a discontinuity in potential or temperature at the interface of contact.
At smaller scales, as discussed in the pioneering works of Holm~\cite{holm2013electric}, the conductivity at contact spots exhibits a continuous change in potential (temperature) for electrical (and thermal) contacts.
In a manner analogous to early models of mechanical contact involving rough surfaces, the true contact area can be represented by a collection of discrete circular contact spots distributed across the nominal contact area, and potentially extending even beyond it~\cite{greenwood1967elastic, greenwood1966constriction}.
The latter case is possible when there is a lack of scale separation between contacting shapes and roughness characteristics, see~\cite{greenwood1984surface,yastrebov2019elastic}.

One of numerous examples that accounts for inelastic deformations in contact is the work of Kogut and Komvopoulos~\cite{kogut2003electrical}, where the authors used a fractal geometry with a simple overlap (cut-off) model to establish a theoretical connection between fractal roughness parameters and constriction resistance, assuming elasto-plastic deformation.
However, it is worth recalling here, that overlap models were shown to produce erroneous results and should not be used for quantitative analysis~\cite{pei2005finite,greenwood2006simplified,dapp2012prl}.
In the elastic regime, Barber~\cite{barber2003bounds} rigorously demonstrated the equivalence between electrical/thermal contact resistance and contact normal stiffness.
Consequently, within the contact mechanics community, this equivalence is often employed as a justification for disregarding a separate study of thermoelectrical diffusion equation in mechanical contact of rough surfaces.

However, in real-life applications, electrical and thermal resistance are not merely reduced to contact stiffness.
First, the contact-induced deformation is often accompanied by inelastic deformations.
In addition, in case of thermal resistance, due to the presence of electrically insulating or weakly conducting oxide films and surface contamination, the conductive contact area is reduced~\cite{holm2013electric},\cite[Ch.1-4]{slade2017electrical}.
Furthermore, in the case of thermal conductivity, additional convective and radiative contributions to heat exchange cause the conductivity to deviate from the strict mathematical equivalence between elastic stiffness and thermal/electric resistance established by Barber~\cite{barber2003bounds}.
Therefore, in all aforementioned contexts, a study of coupled thermo-/electromechanical problems is worth investigation.

The growing demand for micro-electric devices~\cite{tu2003recent} has spurred an increasing need for \textit{electrical} contact models at the microscale.
The constriction resistance model remains consistent with experimental studies~\cite{watanabe1986new} and can be extended to various contact shapes~\cite{nakamura1993constriction, sano1985effect} as well as to multi-spot contact configurations~\cite{thomas1970establishment, nakamura1986computer}.
Contact resistance is influenced by the hardness and resistivity of the contacting solids, but is also affected by oxidation at the contact surface~\cite{cuthrell1973electric}.
However, the study of simple conductive effects reaches its limits at small scales, where electrical resistance must incorporate the ballistic description of electrons' motion, also known as Sharvin's resistance~\cite{mikrajuddin1999size, jensen2005low}.
Similar research has been conducted for \textit{thermal contact resistance}, which has numerous applications in aerospace, automotive and electronic domains, including conductivity of bolted joints and of thermal cooling devices~\cite{kumano1994mechanical}.
This issue is experimentally represented by the thermal contact resistance (TCR), which necessitates careful attention to the set-up precision~\cite{kempers2009high} in both steady-state and transient studies~\cite{fieberg2008determination, burghold2015determination}.
Experimental advancement was accompanied by theoretical~\cite{cooper1969thermal, lambert1997thermal, zou2008fractal, jackson2008multiscale} and numerical investigations~\cite{sadowski2010model, murashov2015numerical} of thermal resistance between rough surfaces in contact.

\begin{figure}[htb!]
\includegraphics[width=1\textwidth]{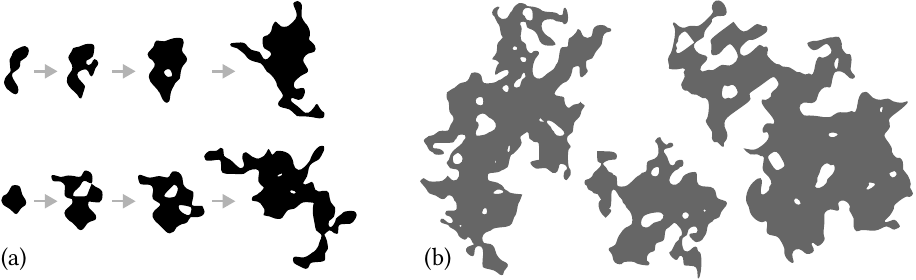}
\caption{\label{fig:0}Simulation results of the true contact-area between self-affine random rough surfaces  taken from~\cite{Yastrebov2015ijss}: (a) evolution of separate contact spots under increasing load, (b) snapshots of separate clusters highlighting the complexity of the shape.}
\end{figure}

The primary motivation for this study stems from the observation of contact spot geometry formed between model rough surfaces (self-affine random geometry).
Figure~\ref{fig:0}(a) illustrates how individual contact spots evolve under increasing load~\cite{Yastrebov2015ijss}, while in (b) three separate contact spots obtained from similar simulations demonstrate high complexity both in terms of connectedness and boundary shape.
Even under relatively small loads, very complex contact spots can be formed if the spectral content of roughness~\cite{nayak1971random} is sufficiently rich.
However, basic models of contact resistance assume that individual contact spots distributed over the nominal contact area possess simple shapes: elliptical or circular.

In contrast, the realistic contact shapes (Fig.~\ref{fig:0}) can be non-simply connected (having holes) and exhibit highly complex boundaries.
This complexity can be characterized by the ratio of the square root of the area to its perimeter, also known as compactness.
In this study, we investigate the impact of connectedness and compactness of individual contact spots on their conductive properties.
Instead of studying contact spots obtained in direct numerical simulations or rough contact, as shown in Fig.~\ref{fig:0}, we construct relatively simple models that capture the primary features of such spots: (1) connectedness, and (2) low compactness.
The first effect is represented by an annulus shape with a varying ratio of inner to outer radius.
Compactness, as a first approximation, is modeled through a flower-shaped spot with varying numbers of petals and their lengths.
Additionally, we explore "multi-petal" configurations to generalize our findings.
The complexity of contact spot geometry is further addressed by modeling self-affine contact spots, akin to multiscale petals, paraphrasing Archard with "petals on top of petals on top of petals." The \emph{primary objectives} of this study are twofold: firstly, to understand the subtle relationship between the geometry of such complex shapes and their thermal and electrical conductivity; and secondly, to construct a simplified model capable of predicting this conductivity based on a set of basic geometrical characteristics.

This paper is structured as follows: Section~\ref{sec:methods} briefly introduces the numerical methods employed in this study.
The subsequent three sections study the effects of the topology and the shape of contact spots on simple examples.
Specifically, Section~\ref{sec:connectivity} explores the influence of non-simple connectedness using the flux through an annular spot as a case study.
Section~\ref{sec:flower} examines the effect of compactness on conductivity of a simplified model of a flower-shaped spot, characterized by equally spaced, uniformly sized petals, and similar formations.
Section~\ref{sec:self-affine} extends this examination to self-affine spots, which could be seen as flower-shaped spots with polydispersed petals following a self-affine distribution.
The paper concludes with Section~\ref{sec:conclusion}.

\section{Methods\label{sec:methods}}

Electrostatics and stationary thermal conductivity are described by elliptic partial differential equation, also known as Poisson's equation $\Delta U = S$, where $U$ takes a role of electrostatic potential or temperature, $S$ is the source/sink term and the flux is given by $\vec j = -k\nabla U$ with $k$ being the electric or thermal conductivity.
Throughout the paper we assume linear isotropic, homogeneous and constant conductivity $k$, i.e.
$\nabla k = 0$, $\partial k/\partial U = 0$.

In this study, to solve Poisson's equation (elliptic PDE) on an isotropic, linear and homogeneous half-space we used an in-house implementation of the boundary element method (BEM)~\cite{bonnet1999boundary}
in its fast-BEM version based on hierarchical matrices~\cite{grasedyck2005adaptive,grasedyck2003construction,hackbusch2015hierarchical,chaillat2017theory}.
The solution of the Poisson's equation can make use of a particular solution, formulated with a Green function $\mathcal G(\vec x,\vec y)$,
resulting in an integral equation involving only the contact area $A$ where a constant potential $U(\vec x) = U_0$ is prescribed:
\begin{equation}
  U(\vec x) = \int_{A} \frac{j_n( \mathbf{y} )}{k} \mathcal{G}( |\mathbf{x} - \mathbf{y} | ) \,dS_y 
  \label{eq:integral1}
\end{equation}
The Green's function depends only on the distance between the "source" point $\vec y$ and the "observation" point $\vec x$ $\mathcal G(|\vec x - \vec y|)$ stands for the elementary solution of the Poisson's equation. 

\begin{figure}
  \centering
  \includegraphics[width=1\linewidth]{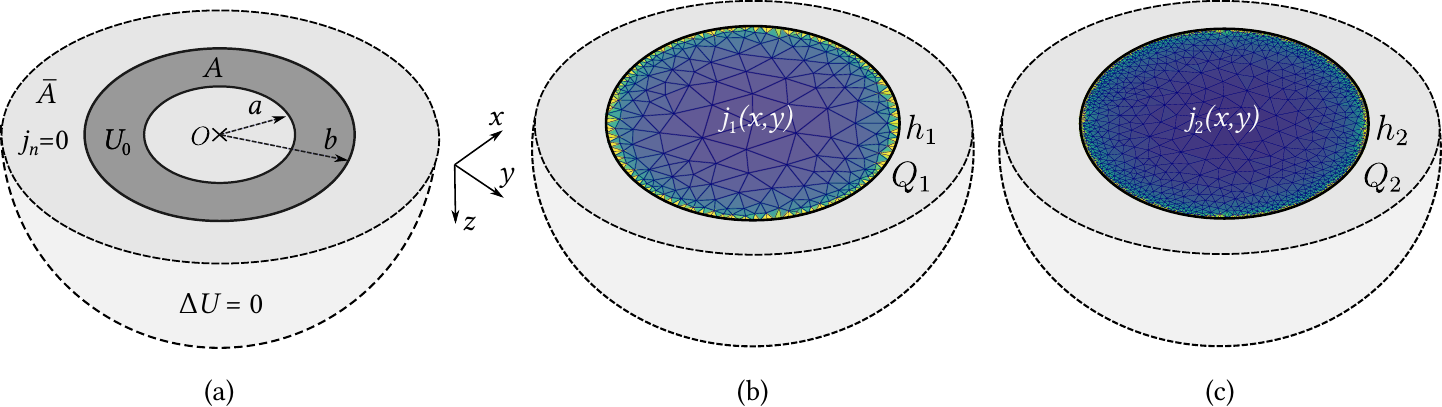}
  \caption{ \label{fig:richardson}Conductivity through a contact spot on the half-space: (a) -- the problem set-up for an annulus with prescribed potential $U_0$ in the zone $A$ and zero flux $j_n=0$ outside, i.e. on $\bar A$; (b-c) -- an illustration of the Richardson extrapolation technique: the same problem is solved on two meshes possessing different reference sizes $h_1,h_2$, the resulting fluxes $Q_1,Q_2$ are extrapolated to a limit value $h\to 0$ assuming a linear convergence with the mesh size $h$ using Eq.~\eqref{eq:richardson}.}
\end{figure}

In this study we focus on a homogeneous Poisson's problem $\Delta U = 0$ with a constant potential $U_0$ prescribed on the contact area $A$ and zero flux $j_n=0$ outside, i.e. on $\bar A$, see Fig.~\ref{fig:richardson}(a).
This conductivity problem is equivalent to a mechanical contact problem of elastic indenting a half-space by a flat indenter with the identical section  $A$~\cite{barber2003bounds}. The only difference is that the flux $j_n$ is replaced by the contact pressure $p$ and the potential $U$ is replaced by the normal displacement $u_z$, which could be readily derived from the Boussinesq's solution~\cite{boussinesq1885application} as
\begin{equation}
  u_z( \mathbf{x}) = \int_{A} \frac{2 (1 - \nu^2) p( \mathbf{y} )}{E} \mathcal{G}( |\mathbf{x} - \mathbf{y} | ) \,dS_y 
  \label{eq:integral_uz}
\end{equation}
Here, we assume that the half-space is made of an isotropic material linearly elastic material with $E$ being the Young's modulus and $\nu$ being the Poisson's ratio. To go beyond linear contact problems, the analogy still persists between the normal contact stiffness, the derivative of the contact force $F$ to normal displacement $u_z$, and the electrical/thermal resistance $R$ as soon as the contact area is identical to the conducting area $A$:
\begin{equation}
  R = \frac{E}{2k(1-\nu^2)} \left[\frac{\partial F}{\partial u_z}\right]^{-1}.
  \label{eq:RandStiffness}
\end{equation}
For a further discussion and a rigorous derivative of this equivalence the reader is referred to~\cite{barber2003bounds}.

The integral equation (Eq.~\ref{eq:integral1}) can be turned into a linear system to solve, involving the construction of a fully-populated matrix.
Hierarchical matrices allow to overcome the drawbacks of the classical BEM, i.e. excessive storage and a tedious resolution of linear system of equations with a full matrix.
The open-source Python code is shared under BSD3 license in~\cite{OurCode} along with the details of its implementation and a short documentation.
Constant interpolation triangular elements were used.
In addition, some problems were also solved using our in-house finite element suite Z-set~\cite{besson1997large,Z-set} on a cylindrical approximating of the half-space with the height and radius of the cylinder set to be much greater than the size of the conducting spot.

To achieve accurate results using the Boundary Element Method (BEM), the problem is addressed using two distinct meshes, each characterized by a relatively fine granularity and differing in the number of nodes.
Results are then extrapolated employing the Richardson extrapolation technique~\cite{richardson1911approximate}, as illustrated in Fig.~\ref{fig:richardson}. 
Two meshes are built, approximating the same contact area, but with different reference mesh size $h_1 > h_2$ (here, we take $h_1 = 2 h_2$). 
Since the flux converges linearly with the mesh size, a simple extrapolation would give~\cite{zienkiewicz2000vol1}:
$$Q^* \approx 2Q(h_2) - Q(2h_2).$$
In a general case of two different reference mesh-sizes, we get
\begin{equation}
    Q^* \approx \frac{h_1 Q(h_2) - h_2 Q(h_1)}{h_1 - h_2}.
    \label{eq:richardson}
\end{equation}
More details and numerical experiments could be found in~\cite{beguin2024phd}.
Since we use linear elements, the accuracy of the geometry approximation is also dependent on the mesh size, therefore the Richardson extrapolation is not exact. Nevertheless, for the geometrical error, the ratio of the difference between approximate and the true area to the true area scales as $\approx h^2/(8R^2)$ where $R$ is the curvature radius, therefore we ignore this error.

For the FEM, in addition to mesh-extrapolation we need to introduce a correction related to the finite size geometry, namely, the dependence on the ratio of the average spot radius to the size of the simulated domain, which can be seen as a geometrically related small parameter $\epsilon_g = b/R$ (see Fig.~\ref{fig:fig_mesh_hole}). The primary, mesh-related, small parameter is given by $\epsilon_h = h/b$. Then, if we assume a linear dependence on $\epsilon_g$, we can write first terms of the expansion around the exact solution $Q^*$ as:
$$
 Q(\epsilon_h,\epsilon_g) = Q^* + c_h \epsilon_h + c_g \epsilon_g + o(\epsilon_h,\epsilon_g).
$$
In this case, three simulations are needed to determine the value of $Q^*$ and coefficients $c_h$ and $c_g$ in which two values of $\epsilon_h$ and of $\epsilon_g$ should be used in any combination. More details on the mesh and geometry corrections could be found in~\cite{beguin2024phd}.
 
\section{Conductivity of a not simply connected spot\label{sec:connectivity}}

The total flux of a single circular spot of radius $a$ between two half-spaces made of the material with the same thermal/electric conductivity $k$ was given in~\cite{carslaw1947conduction}: 
\begin{equation}
    Q_{\circ} = 4 k a U_0,
    \label{eq:Q_circ}
\end{equation}
where $U_0$ is the difference between electric potential or temperature between the spot and an infinitely remote boundary.

It assumes stationary conductivity and perfect insulation outside the circular spot, i.e.
we ignore convective and radiative heat exchange for the thermal problem and ignore tunnel effect or electrical breakdown for the electric problem.
Notably, the total flux is proportional to the radius of the spot.
The flux distribution within the spot is axially symmetric and thus could be expressed in polar coordinates as a function of radius $r$:
\begin{equation}
    j^{\circ}_n(r) = \vec j\cdot\vec n = \frac{ 2 U_0 k }{ \pi \sqrt{ a^2 - r^2 } },
    \label{eq:jn_ana}
\end{equation}
where $\vec n$ is the outer normal, see~\cite{carslaw1947conduction}.
The flux diverges as $1/\sqrt{\xi}$ near the boundary where $\xi$ is the distance to this boundary.
This solution is equivalent to the contact pressure of a circular stamp pressed in an elastic half space~\cite{sneddon1995fourier} and to the normal stress distribution of an external circular crack~\cite[p.
377]{tada2000handbook}.
An analytical solution for conductivity of an elliptic spot on a half-space was later obtained in~\cite{holm2013electric}.

\begin{figure}[ht!]
    \centering
    \resizebox{1\textwidth}{!}{\input{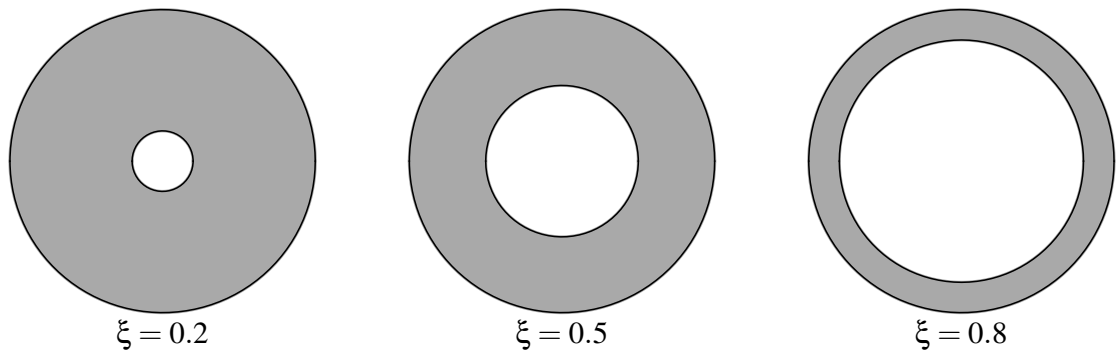}}
    \caption{\label{fig:fig_hole}
    Geometry of the annular contact spot for various ratios of the internal to external radii $\xi = \{0.2, 0.5, 0.8 \}$.}
\end{figure}

In this section, we study the total flux through an annulus spot of different internal radius.
The question is to which extent the flux is altered by the presence of small internal holes in not simply connected spots.
The internal radius is $r=b$, the external one $r=a$, their ratio is denoted by $\xi=b/a \in [0,1)$.
The boundary conditions remain the same as for the circular spot: the potential is set constant $U=U_0$ at the annulus ($b\le r\le a$) the zero flux is set elsewhere ($r>a$ and $r<b$), at infinity the potential is set to $U=0$.
The geometry is shown in Fig.~\ref{fig:fig_hole} for three different values of $\xi$.
We are mainly interested in the asymptotic evolution of the total flux $Q(\xi)$ in the limit of $\xi \ll 1$.
The flux diverges at the two borders of the annulus, however the total flux $Q_a$ should be continuous and decreasing with respect to the relative hole radius $\xi$.

As $\xi\to0$, the total flux should tend to $Q_{\circ}$, whereas for $\xi\to1$, the flux should vanish $Q_a\to0$.
We could also conjecture, from a physical nature of the phenomenon, that the flux should be a concave function of $\xi$ with zero derivative at $\xi=0$, meaning that small holes should not affect considerably the flux.
We could also conjecture that the local singularity on the inner edge should be less pronounced than on the outer edge being regularized to some extent by the interaction with the whole inner border especially for small size holes.

Smythe~\cite{smythe1951capacitance} was the first to solve this problem using a superposition method and provided methods to approximately evaluate the flux along with few tabulated results for the annulus conductivity.
Other authors~\cite{cooke1963some,collins1963solution,fabrikant1993dirichlet} expanded Smythe's work by reformulating the problem as a triple integral or as a Fredholm integral equation of second kind.
However, those solutions do not provide closed form formulas for the flux; the influence of the effect of the hole cannot be easily deduced.
Fabrikant~\cite{fabrikant1993dirichlet} proposed an iterative method for the resolution of the integral system.
An alternative solution was obtained by Love~\cite{love1976Inequalities} based on constructing upper and lower limits series provided the first terms as:
\begin{equation}
    \frac{Q_{\text{\tiny Love}}}{Q_{\circ}} = 1 - \frac{4}{3 \pi^2} \xi^3 - \frac{8}{15 \pi^2} \xi^5   
    - \frac{16}{27 \pi^4} \xi^6 - \frac{92}{315 \pi^2} \xi^{7} - \frac{416}{675\pi^4} \xi^8  + o(\xi^8)
    \label{eq:Qlove}
\end{equation}

\begin{figure}[ht!]
  \centering 
  \includegraphics[width=0.8\linewidth]{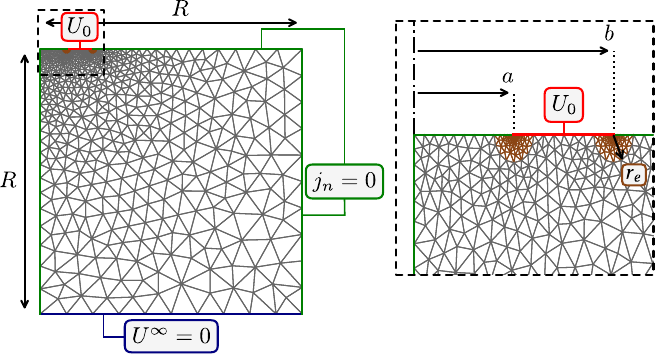}
  \caption{ \label{fig:fig_mesh_hole} Mesh definition and boundary conditions for modeling of an annulus contact
  spot, using the parameters $ \xi = 0.5 $, $h_{\min} / r_0 = 0.005$ and $R/a = 5$}
\end{figure}

We used the finite element method (FEM) to solve this problem numerically as an axisymmetric problem.
The mesh has to be refined near the annulus edges to capture singularities.
To do so, two semicircular insertions of radius $r_e$ are constructed near the singularity points (see Fig.~\ref{fig:fig_mesh_hole}). 
The element size at the annulus's edge is set to $h_{\min}$ such that $h_{\min} / b \ll 1$.
Far from the annulus, the mesh size is set to be coarser $h_{\max}$, as $h_{\max} \gg h_{\min}$.
Supplementing the mesh refinement near the annulus' edge, the mesh size is set to $h_e$, at the edge of the semicircular object setting $h_e = 10 h_{\min}$.

The geometry of the axially symmetric problem is restricted to a square plane of length and width $R \gg b$.
The mesh size on the outer edge must be much smaller compared to $b$, i.e.
$h \ll b$ and near the internal edge $h\ll a$.
However, far from singularities the mesh could be coarser.

\begin{figure}[ht!]
    \centering
    \resizebox{1\textwidth}{!}{\input{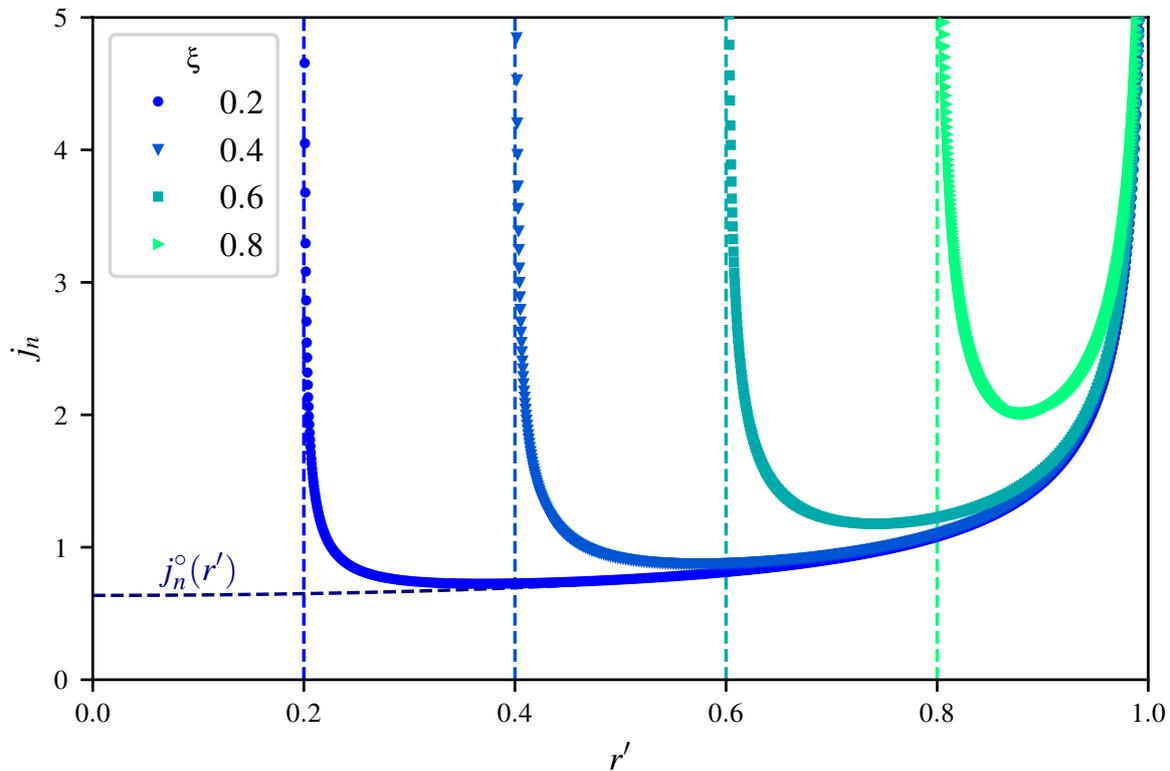} }%
    \caption{ \label{fig:j_hole} Finite element results of the normal flux distribution for annulus spot for different values of $k$, dashed line represents the normal flux of a circular spot.
}
\end{figure}

The finite element results for the flux distribution for several values of $\xi$ are shown in Fig.~\ref{fig:j_hole} and compared with the analytical solution for the circular spot~\eqref{eq:jn_ana}.
The density of result points allows the reader to judge on the density of mesh used for this solution (the finite element mesh is provided in Supplementary material~\cite{supplementary}).
As expected, the normal flux diverges at both edges and the singularity at the hole (internal edge) is less pronounced than on the outer edge: it decreases faster than the one on the outer edge.
Nakamura~\cite{nakamura1993constriction} also studied this problem, by both FEM and BEM to assess the BEM accuracy with respect to the FEM resolution, however, the author ignored the existing analytical solution.

\begin{figure}[ht!]
 	\centering
    \resizebox{1\textwidth}{!}{\input{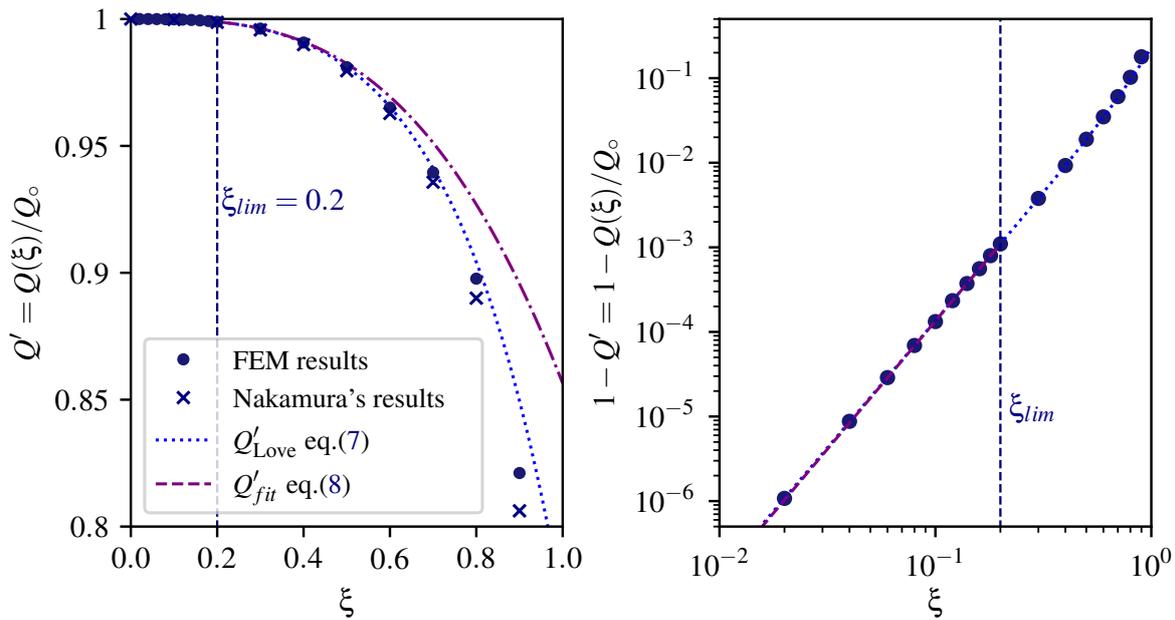} }
    \caption{\label{fig:Q_hole}Total flux of an annulus as a function of internal to external radius ratio $\xi$: finite element results (circles), the least squares fit of an offset power-law for the asymptotic solution $Q'_{\text{fit}}$ for $\xi<0.2$ and an approximate analytical solution by~\cite{love1976Inequalities}
    $Q_{\text{\tiny Love}}/Q_{\circ}$.
    (a) -- normalized total flux $Q'(k) = Q(k)/Q_{\circ}$; (b) -- normalized flux difference compared to the circular flux $(Q_{\circ} - Q(k))/Q_{\circ}$ in log-log scale highlighting the power-law flux evolution.
}
\end{figure}

Simulation results for the \textit{total flux} are shown in Fig.~\ref{fig:Q_hole} with the flux normalized by the one of a circular spot $Q_{\circ}$ (see Eq.~\eqref{eq:Q_circ}).
To identify numerically an asymptotic solution for small holes, we consider a contribution from the hole with a power-law of $\xi$:
\begin{equation}
    Q_{\text{fit}} = Q_{\circ}\left(1- \alpha k^\beta\right),
    \label{eq:Qhole}
\end{equation}
whose parameters $\alpha,\beta$ were identified by least squares method as $\alpha \approx 0.1435$ and $\beta \approx 3.028$ in the interval $\xi<\xi_{\lim}=0.2$, which are very close to analytical results of~\cite{love1976Inequalities,fabrikant1993dirichlet}, $\alpha = 4/(3\pi^2) \approx 0.1351$ and $\beta=3$ (see Eq.~\eqref{eq:Qlove}).
Nakamura's~\cite{nakamura1993constriction} results calculated by the BEM are also displayed for comparison, however, because of the lack of Richardson type extrapolation and convergence study, they underestimate the flux value.
The Love's solution slightly overestimates the flux for higher values of $\xi$, but this could be readily improved by including a larger number of terms in his series.
In fact, all polynomial coefficients of $\xi$ for the series expansion of the flux function are negative and adding new terms will slightly reduce the flux.
Nevertheless, the first terms in Eq.~\eqref{eq:Qlove} are in very good agreement with the numerical results at least for small values of $\xi$.

Expectedly, we conclude that the total flux is very weakly dependent on the presence of small holes in annulus spots because the corrective term is of order $\xi^3$ with a small factor $\sim0.1$.
Therefore, we could conclude that not simply connected spots, at least for holes located far from the outer boundary, conduct almost as well as simply connected spots with the same outer boundary.
In addition to this axisymmetric study, one could conduct a similar study but with a hole placed with some eccentricity with respect to the center.
Such a study would provide an even stronger argument on the effect of such holes in conductivity problems, however, this study is not included in the scope of the current paper.

\section{Conductivity of flower-shaped spots\label{sec:flower}}

\begin{figure}[ht!]
    \centering
    \resizebox{1\textwidth}{!}{\input{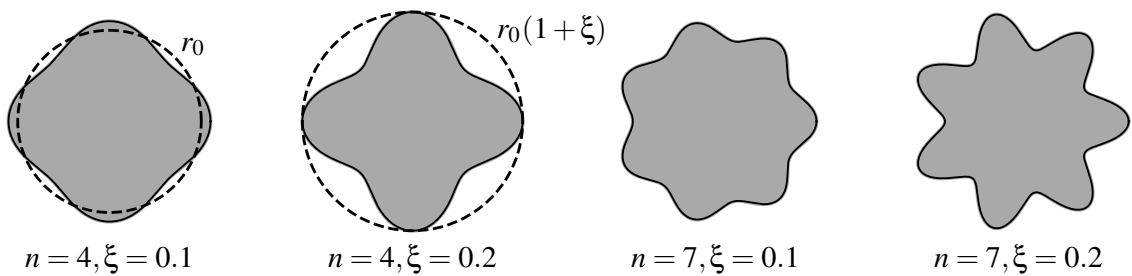}}
    \caption{\label{fig:flower_ex}Examples of flower-shaped spots with $\xi = \{0.1,0.2\}$ and the number of petals $n = \{4, 7\}$.
}
\end{figure}

To mimic complex shapes formed by contact between random rough surface, we first consider a simple geometrical model which we call \emph{flower-shaped} spot whose boundary is described by the following equation in polar coordinates
\begin{equation}
    r(\theta) = r_0 + r_1 \cos(n \theta ) = r_0 \left(1 + \xi \cos(n \theta )\right),\quad r_1 < r_0
    \label{eq:r_flower}
\end{equation}
where $r_0$ is the mean radius, $r_1$ is the half-length of petals and $n$ is their number.
So the two positive dimensionless parameters describing the shape are $\xi=r_1/r_0 < 1$  and $n$.
Different flower-shaped spots are presented in the Fig.~\ref{fig:flower_ex}.
Note that the average radius does not change with the number of petals nor with their length and is equal to $\langle r\rangle=r_0$.
Circles of radius $r_0$ and $r_0(1+\xi) = r_0 + r_1$ are also shown in the figure; corresponding to minimal and maximal limits for the resulting flux $Q_{\min}(r_0) = Q_{\circ} = 4kr_0U_0$, $Q_{\text{\tiny up}} = Q_{\circ}(r_0+r_1) = 4kr_0(1+\xi)U_0$,
\[
1 \le \frac{Q(r_0,\xi,n)}{Q_{\min}} \le 1+\xi
\]

The particularity of these flower-shaped spots is that the perimeter increases with the number of petals whereas the area remains constant.
They are given by the following equations
\begin{subequations}
\label{eq:AP_flower}
\begin{align}
    A &= \pi r_0^2 \left( 1 + \frac{\xi^2}{2} \right), \label{eq:A_flower} \\
    P &= r_0 \int\limits_{0}^{2 \pi} \sqrt{1 + \left( n \xi \sin( \theta ) \right) ^2} d \theta, \label{eq:P_flower}
\end{align}
\end{subequations}
We can notice that the integrand depends on the dimensionless parameter $n'=n\xi \in \mathbb R$ which can take arbitrary real values contrary to the integer number of petals $n\in\mathbb N$.
To characterize the shape of the flower-shaped spot, we could also use compactness $C$ being the ratio of the square root of the area to the perimeter:
\begin{equation}
    C(n') = \frac{ \sqrt{A} }{P} = \sqrt{\pi}\frac{\sqrt{1+\xi^2/2}}{4E(in')},
    \label{eq:C_flower}
\end{equation}
where $E(x) = \int_{0}^{\pi/2} \sqrt{1 -  x^2\sin^2(\theta) } d \theta$ is the complete elliptic integral of the second kind and $i$ is imaginary unit.
For small values of $\xi$, we could assume that the compactness depends on $n'$ only, for $n'\to0$, $E(in')\to\pi/2$ and the compactness tends to the maximal value, i.e.
the compactness of a circle $C \to C_{\circ} = 1 / (2 \sqrt{ \pi } ) \approx 0.282$.

\subsection{Conductivity results}

This study was conducted using both FEM and BEM.
Despite the fact that the BEM is more appropriate for half-space approximation, the comparison of two methods with an extrapolation in terms of the size of simulated domain employed in the FEM is also used to assess the validity of the implemented BEM.
The flower shape presents a dihedral symmetry $D_n$ which allows us to use only one mesh section (half-petal) with symmetric boundary conditions imposed on the lateral sides out of $2n$ sections needed to construct the full spot for the FEM.
The same size reduction can be performed with the BEM using instead the repeatability of the solution.
With an increasing number of petals, the angle of the central element near the axis of symmetry sharpens. 
Consequently, to prevent any deterioration in the solution quality, the smallest simulated angular sector was configured to be approximately \( \pi / 6 \). 
This setup necessitates simulating a larger segment than what is required by symmetry considerations only.

\begin{figure}[ht!]
    \centering
    \includegraphics[width=1\textwidth]{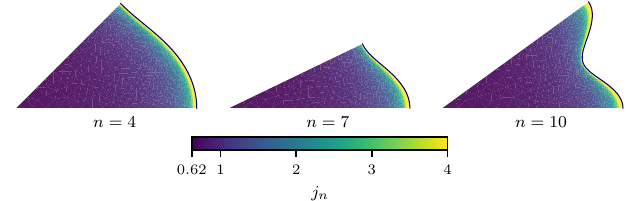}
    \caption{\label{fig:jn_flower}Simulation results for the normal flux for flower-shaped spots with $n = {4, 7, 10}$ and $\xi = 0.1$.}
\end{figure}

Simulation results of the normal flux are presented in Fig.~\ref{fig:jn_flower} for $n=\{4,7,10\}$ petals and for $\xi = 0.1$.
For the first two cases, the symmetry is fully exploited whereas for $n=10$ the sector's angle is set to $ \pi / 5$ to preserve good mesh quality in the center.
Meshes are refined near the outer edges where the flux is singular.
For  $n=4$ the total flux is $Q\approx 1.0084 Q_{\min}$, for $n=7$ the total flux is $Q\approx 1.0150 Q_{\min}$, and for 
$n=10$ the total flux is $Q\approx 1.0209Q_{\min}$.
So there is a trend to increase the total flux with the increasing number of petals.
Visually we can also observe that the singularity in the trough (petal's root) is weaker than the one near the crest (petal's extremity).
The more petals we have, the weaker the flux intensity in the trough because of the increasing interaction with the neighboring petals; a similar trend was observed for the annular spot for small internal radii.

\begin{figure}[ht!]
    \centering    
    \resizebox{1\textwidth}{!}{\input{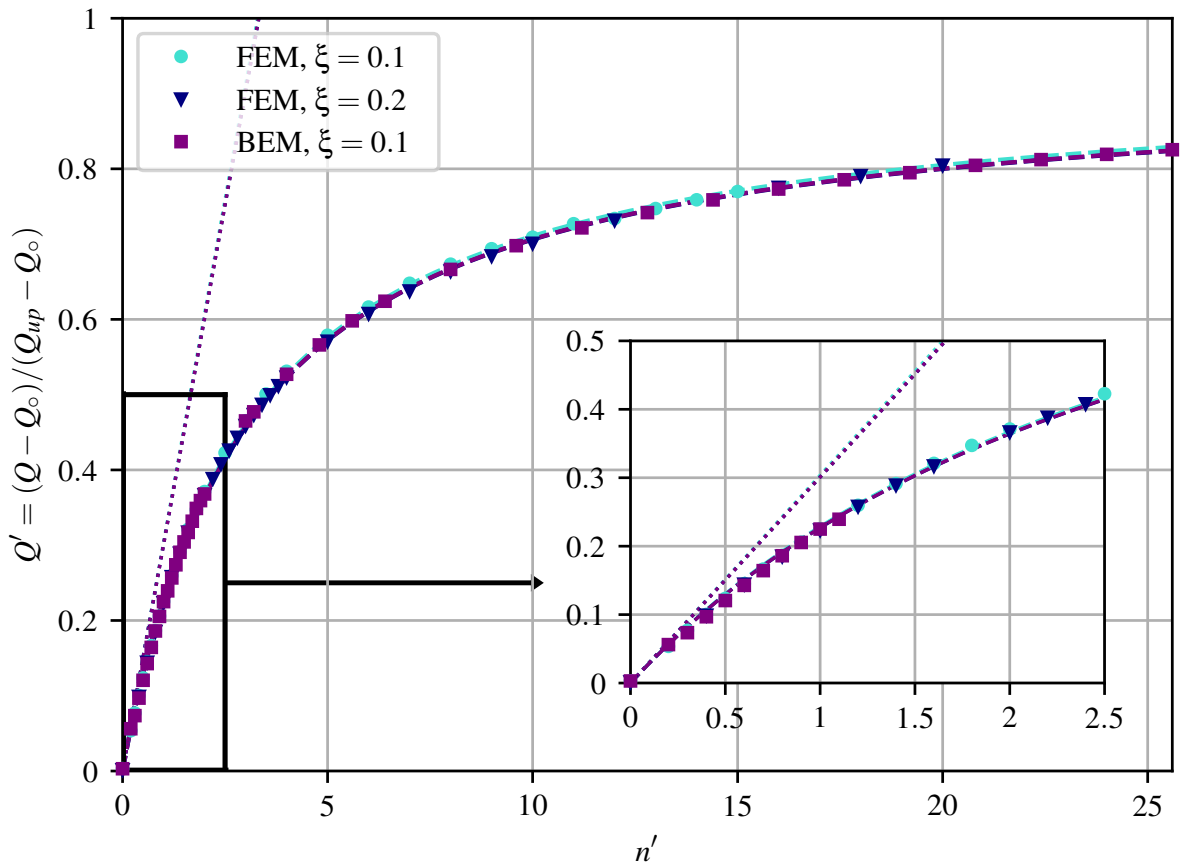}}
    \caption{\label{fig:Q_flower}Simulation results for the normalized total conductivity for the flower-shaped spot as a function of normalized petal's parameter $n'$ for different $\xi$, three independent least squares fit of function~\eqref{eq:Q_fit} are also plotted; the three corresponding tangents at the origin are also plotted.
}
\end{figure}

In total, $97$ simulations\footnote{Without counting simulations used to employ Richardson extrapolation.} were carried out for $\xi=0.1$ and $n\in (1,256)$ as well as for $\xi=0.2$ and $n\in(1, 100)$.
The resulting total flux with respect to the normalized number of petals $n'$ is presented in Fig.~\ref{fig:Q_flower}.
Those results are computed with Richardson extrapolation obtained by two meshes of different density for FEM and BEM and using a similar extrapolation for domain's dimensions in case of FEM simulations.

The total flux, offset by the minimal flux $Q-Q_\circ$ could be normalized by the difference between the maximal and the minimal theoretical fluxes, corresponding to circular spots of radii $r_0(1+\xi)$ and $r_0$, respectively, i.e.
we get in the denominator $Q_{\text{\tiny up}} - Q_{\circ} = 4 k r_0 \xi U_{0} = \xi Q_\circ$.
This normalization results in a universal curve for the total normalized flux for any $r_0$ and $\xi$:
\begin{equation}
    Q' = \frac{Q - Q_{\circ}}{Q_{\text{\tiny up}} - Q_{\circ}} = \frac{Q - Q_{\circ}}{\xi Q_{\circ}}
    \label{eq:Q_flower_norm}
\end{equation}
The normalized flux evolution seems to be logarithmic, but from the physical point of view, the flux cannot overpass $Q_{\text{\tiny up}}$, therefore we shall require that the bounds $0\le Q' \le 1$ are respected for all $n'$.
The resulting flux is well fitted by a two-parameter function which, however, was found empirically:  
\begin{equation}
    Q'_{\mathrm{fit}}(n') = a \left( 1 - \frac{1}{b n' + 1 } \right), \quad 0 < a \le 1.
    \label{eq:Q_fit}
\end{equation}
This fit function is plotted in Fig.~\ref{fig:Q_flower} along with FEM and BEM simulation results.
The coefficients determined by least squares fit are presented in Table~\ref{tab:Coef_flower}.
The slope at $n' = 0+$ is equal to the product $a b$.
Even if the coefficients $a$ and $b$ are slightly different for different sets, this slope remains close for all three independent fits, and roughly equals to $0.3$.
Combining Eqs.~\eqref{eq:Q_flower_norm} and \eqref{eq:Q_fit}, we obtain the following phenomenological equation for the total flow of a flower-shaped spot:
\begin{equation}
 Q =  Q_{\circ}\left(1 + a \xi \left( 1 - \frac{1}{b n \xi + 1 } \right)\right) \approx
 Q_{\circ}\left(1 + 0.923 \xi \left( 1 - \frac{1}{0.326 n \xi + 1 } \right)\right),
\end{equation}
where $Q_\circ = 4k r_0 U_0$.
For the infinite number of petals $n$ of finite half-length factor $\xi$, there is a limit flux given by this fit, $ \lim_{n' \to \infty} (Q'_{\mathrm{fit}} ) = a \approx 0.923$ and this limit is independent of $\xi$.
However, the validity of the suggested fit beyond the studied interval of $n'$ cannot be taken for granted. An argument in favor of such a limit $a<1$, i.e. that the total flux for the infinite number of petals remains below the flux of a circular shape of radius $r_0(1+\xi)$, could me made based on the area of the conductive spot. Indeed, the area of the full circular spot is considerably bigger than this of the flower-shaped spot $\pi r_0^2(1+\xi)^2 > \pi r_0^2 (1+\xi^2/2)$. On the other hand, the small features (infinitely thin petals) should not strongly affect the conductivity of the spot, thus suggesting that possibly the flux should simply tend to $1$... However, the current fit function could not be properly approximate the data if one sets $a=1$. The question of a rigorous definition of the limit flux for the infinite number of petals \emph{remains open}. But our initial guess, for the fitted parameters is given by the following limit flux:
\begin{equation}
\lim\limits_{n\xi\to\infty} Q \approx Q_\circ(1+ 0.923 \xi) < Q_{\text{\tiny up}} = Q_\circ(1+\xi).
\end{equation}

Even though such a flower-shaped geometry, especially in the limit of infinite number of petals, is not very relevant to contact problems between isotropic surfaces, which was at the origin for this study, this limit value presents an interesting by-product of this study.
Among other results one can deduce a relation between $n$ and $\xi$ which ensures $x$ fraction conductivity in the interval $Q_\circ$ and $Q_{\text{\tiny up}} = (1+\xi)Q_\circ$, with $Q=Q_\circ$ for $x=0$ and $Q=Q_\circ(1+\xi/2)$ for $x =0.5$:
\begin{equation}
    x = a \left( 1 - \frac{1}{b n \xi + 1 } \right) \quad \Leftrightarrow \quad  
    n' = n\xi   = \left[\frac{x}{b(a-x)}\right] \quad \Rightarrow \quad n\xi \approx \left[\frac{x}{0.326(0.923-x)}\right],
\end{equation}
Therefore, to reach the mean flux between two limits, i.e.
for $x=0.5$, one would need a spot with $n\xi \approx 3.61$, i.e.
for $\xi=0.2$ one would need approximately $18$ petals and for $\xi=0.1$ a double of that.
However, to reach 75\% ($x=0.75$), for $\xi=0.1$, one would need a spot with approximately $130$ petals.

\begin{table}[ht!]
\centering
\begin{tabular}{c c c c c c}
 			& \multicolumn{2}{c}{Parameters} & \multicolumn{3}{c}{Coefficients}\\
    Simulation & $\xi$ & $n \in$ & $a$ & $b$ & $ab$ \\
    \hline\\[-10pt]
    FEM & $0.1$ & $[1,150]$ & 0.928 & 0.327 & 0.304 \\
    FEM & $0.2$ & $[1,100]$ &  0.923 & 0.326 & 0.301 \\
    BEM & $0.1$ & $[1,256]$ &  0.923 & 0.326 & 0.301 \\
\end{tabular}
\caption{\label{tab:Coef_flower}Least squares fit for coefficients of Eq.~\eqref{eq:Q_fit} for the sets of results of flower-shaped spot obtained using FEM and BEM simulations.}
\end{table}

\subsection{Alternative "multi-petal" spots}

\begin{figure}[ht!]
    \centering
    \resizebox{1\textwidth}{!}{\input{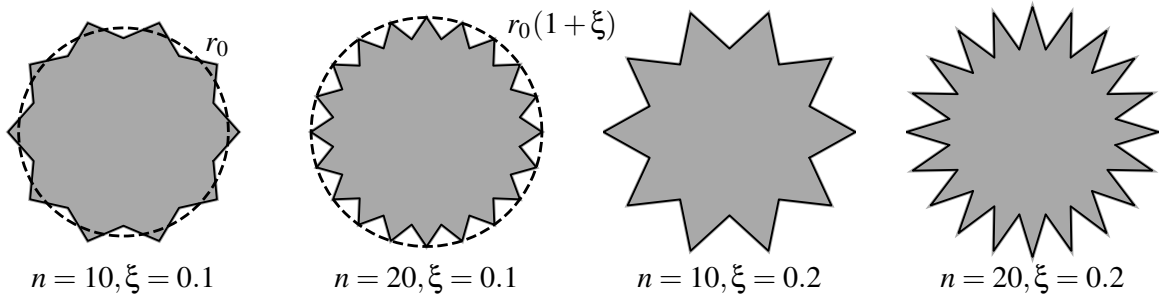}}
    \caption{\label{fig:fig_40_star} Examples of star-shaped spots with $n \in \{ 10, 20\}$ petals, and half-petal length defined by $ \xi \in \{0.1, 0.2\}$.
    }
\end{figure}

The same conductivity study could be conducted on other simple forms possessing a single-scale petal-like structures with the same symmetry properties. Specifically we identified the following shapes: star-shaped and gear-shaped spots shown in Fig.~\ref{fig:fig_40_star} and Fig.~\ref{fig:fig_40_gear}, respectively.
For "stars", each petal is made up by straight lines connecting the roots and extremities of "petals", i.e.
points with radial coordinates $r_0(1 - \xi)$, $r_0(1 + \xi)$.
The number of "petals" (or "rays") as previously is denoted by $n$, and the half-petal length $r_1$ is again determined by the ratio $ r_1 = \xi r_0$.
The gear-shaped spots are made of circular arcs with constant $r=r_0(1-\xi)$ and $r=r_0(1+\xi)$ over equal angular segments.
Contrary to $C^\infty$ flower-shaped spots, star-shaped ones are only of class $C^0$ with respect to $\theta$ whereas gear-shaped spots represent multivalued mapping, so they are not even injective even for a single "petal" or "tooth".

\begin{figure}[ht!]
    \centering
    \resizebox{1\textwidth}{!}{\input{fig_40_gear.pgf}}
    \caption{\label{fig:fig_40_gear} Examples of gear-shaped spots with $n \in \{ 10, 20 \}$ petals, and half-petal lenght defined by $ \xi \in \{0.1, 0.2 \}$.
    }
\end{figure}

The area of these shapes can be readily expressed by an elementary sum of triangles and a regular polygon for the stars, or circular sectors for the gears.
Similarly, the perimeters are easy to find, and the resulting compactness can be also readily computed:
\begin{equation}
\begin{aligned}
    P_{\text{\tiny star}} &= 2 \sqrt{2} n r_0 \sqrt{ 1 + \xi^2 - ( 1 - \xi^2 ) \cos( \pi / n ) },\\[3pt]
    A_{\text{\tiny star}} &=  n r_0^2 ( 1 - \xi^2 ) \sin( \pi / n),\\[3pt]
    C_{\text{\tiny star}} &= \frac{ \sqrt{( 1 - \xi^2 ) \sin( \pi / n)}}{ 2 \sqrt{2n} \sqrt{ 1 + \xi^2 - ( 1 - \xi^2 ) \cos( \pi / n) } }.\\
\end{aligned}
    \label{eq:Star}
\end{equation}

\begin{equation}
    \begin{aligned}
        P_{\text{\tiny gear}} &= 2 \pi r_0 ( 1 + 2 n' / \pi ),\\[3pt]
    A_{\text{\tiny gear}} &= \pi r_0^2 (1 + \xi^2 ),\\[3pt]
    C_{\text{\tiny gear}} &= \frac{\sqrt{ ( 1 + \xi^2 )}}{2 \sqrt{\pi} ( 1 + 2 n' / \pi )}.
\end{aligned}
\label{eq:Gear}
\end{equation}

All those geometric features are summarized in appendix~\ref{app:apc} and in Fig.~\ref{fig:apc}. In summary, the star-shaped area converges to
\[
    A_{\text{\tiny star}} \xrightarrow[]{n\to\infty}  \pi r_0^2 ( 1 - \xi^2 ),
\]
but for all $n$ it is always smaller than the area of gear- and flower-shaped spots:
\[ A_{\text{\tiny star}} < A_{\text{\tiny flower}} < A_{\text{\tiny gear}}.\] 
The gear-shaped spots also have a bigger perimeter than the one of flower and star for a given number of petals:
\[ P_{\text{\tiny flower}} \approx P_{\text{\tiny star}} < P_{\text{\tiny gear}}.\]
Finally, the gear-shaped spots appear to be the least compact shape, while the flower shape is as compact as the star
\[ C_{\text{\tiny gear}} < C_{\text{\tiny flower}} \approx C_{\text{\tiny star}}.\]

\begin{figure}[ht!]
    \centering
    \includegraphics[width=1\textwidth]{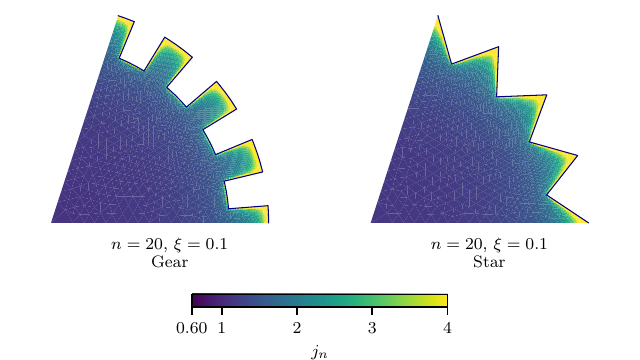}
    \caption{\label{fig:J_mesh_star_gear} BEM result of the flux through a gear- and star-shaped spots of $20$ petals and $\xi = 0.1$.}
\end{figure}

\begin{figure}[ht!]
    \centering
    \resizebox{1\textwidth}{!}{\input{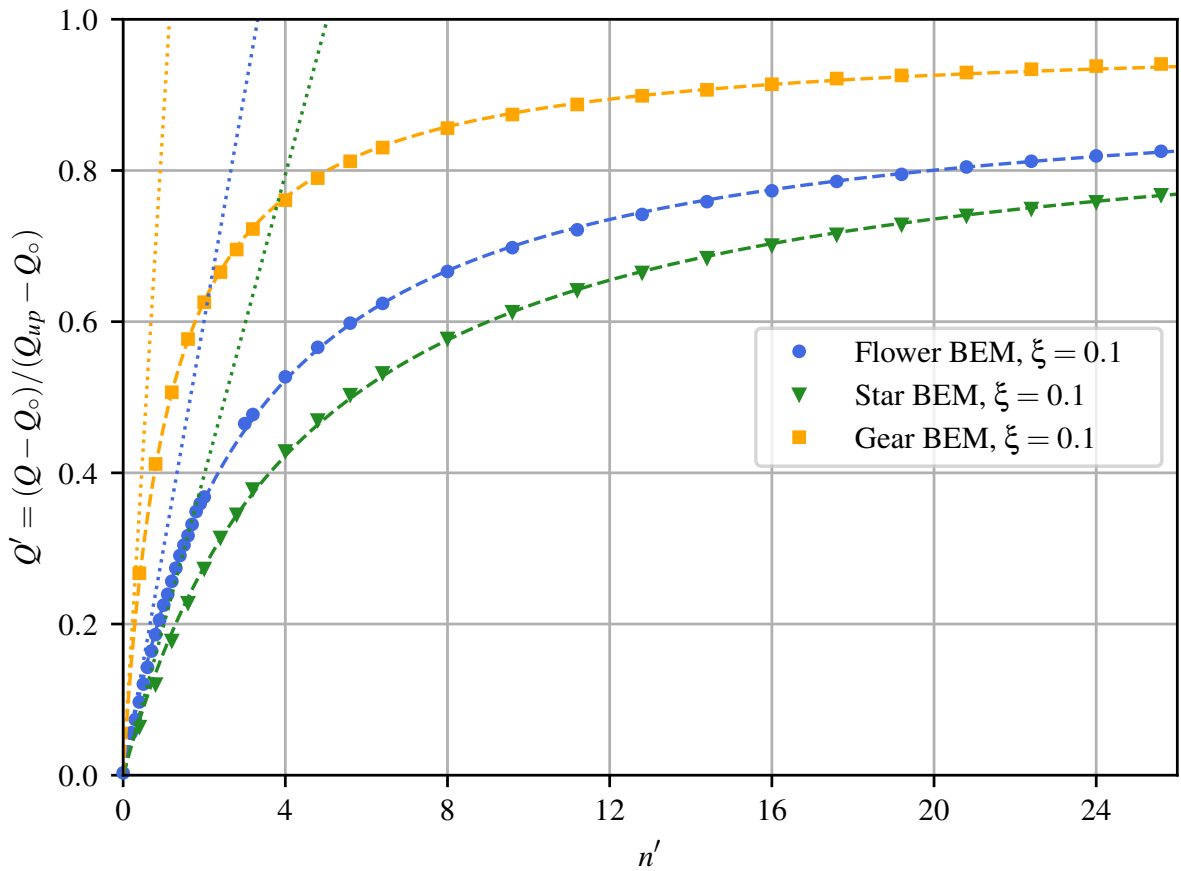}}
    \caption{ \label{fig:Qn_star_gear} Results of the normalized flux for star-, gear- and flower-shaped spots with respect to the normalized "petal" parameter $n' = \xi n$.
Least squares fit of Eq.~\eqref{eq:Q_fit} is also shown.
    }
\end{figure}

The conductive simulations for these shapes were conducted using the fast-BEM.
The star- and gear-shaped spots present the same dihedral symmetry $D_n$ as for the flower-shaped, which allows us again to reduce the problem size to some extent.
Example of simulation results are presented in Fig.~\ref{fig:J_mesh_star_gear}.
In total, $48$ simulations were performed for both gear- and star-shaped spots, for $\xi = 0.1$, and $n \in (4, 256)$.
As previously, the conductive property is assessed by computing the overall flux and helped by the Richardson extrapolation.
Those are again normalized according to~\eqref{eq:Q_flower_norm}, and finally presented in Fig.~\ref{fig:Qn_star_gear} complemented with previous results for the flower-shaped spots. The least square fitted parameters for the same Eq.~\eqref{eq:Q_fit} are shown in 
Table~\ref{tab:param_star_gear_flow}.

Qualitatively, all three types of shapes show the same trend in the total flux evolution with the number of "petals": an initial steep increase and further saturation to a more or less constant value. The thermal conductivity of the gears is higher than that of the flowers, while the star-shaped configuration displays a lower conductivity.
We could attribute this ordering to the only basic geometric parameter which significantly differs for all three of them, namely the area. Another consideration is the amount of area located closer to the outward boundary, which of course is higher for the gear like geometry than those of flower and star.
In the limit of the infinite number of petals the following result (initial guess obtained by extrapolation) is obtained (see parameter $a$ in Table~\ref{tab:param_star_gear_flow}):
\[
    Q^{\lim}_{\text{\tiny star}} \approx 0.903 \quad<\quad Q^{\lim}_{\text{\tiny flower}} = 0.923 \quad<\quad Q^{\lim}_{\text{\tiny gear}} = 0.978.
\]
However, we would like to highlight once again that these values must be seen as a first guess, and a more rigorous assessment (e.g., using an accurate asymptotic analysis) is needed.

\begin{table}[ht!]
\centering
\begin{tabular}{c c c c c c}
 			& \multicolumn{2}{c}{Parameters} & \multicolumn{3}{c}{Coefficients}\\
    Simulation & $\xi$ & $n \in$ & $a$ & $b$ & $ab$ \\
    \hline\\[-10pt]
    Gear  & $0.1$ & $[4, 256]$ &  0.978 & 0.894 & 0.875 \\
    Flower  & $0.1$ & $[1,256]$ &  0.923 & 0.326 & 0.301 \\
    Star  & $0.1$ & $[4, 256]$ & 0.903 & 0.220 & 0.199 \\
\end{tabular}
\caption{
\label{tab:param_star_gear_flow}
Fit parameters for Eq.~\eqref{eq:Q_fit} for the total flux of different multi-petal shapes.
}
\end{table}

\section{Conductivity of self-affine spots\label{sec:self-affine}}

\subsection{Geometry of spots}

Being inspired by shapes of contact clusters occurring in contact of random rough surfaces (see Fig.~\ref{fig:0}), in this section we study contact spots of model complex shapes presenting some randomness. The shapes under study (see Fig.~\ref{fig:ex_ink}) could also recall coffee or ink stains.
To take up the Archard's image of "protuberances on protuberances on protuberances"~\cite{archard1957elastic}, we constructed contact-spots in a self-affine fashion by summing-up multiple harmonics.
The goal is to imitate to some extent realistic contact spots occurring for surfaces with a rich spectral content and to expand the results obtained for single-harmonic flower-shaped spots to more complex forms.

The first step to generate a spot with self-affine boundary is to introduce a periodic perturbation function $h(\theta)$ as a superposition of cosines which individually could be seen as flower-shaped spots:
\begin{subequations}
    \label{eq:rh_ink}
    \begin{align}
        h(\theta) &= \sum_{k=k_l}^{k_s} \xi_k \cos{(k\theta+\theta^0_k)}, \quad h(\theta) = h(\theta+2\pi), \quad \langle h\rangle = 0 \label{eq:h_ink} \\
        \xi_k &= \xi \left(\frac{k}{k_l}\right)^{-(0.5+H)}. \label{eq:r_k}
    \end{align}
\end{subequations}
The summed up harmonics include all integer modes from a fixed interval $k\in(k_l,k_s)$ with amplitudes $\xi_k$ which decay as a strict power-law of the mode number with an exponent involving the Hurst exponent $H \in (0, 1)$ ensuring a self-affinity of the boundary.
The randomness is provided exclusively by the phases $\theta^0_k$ which follow a uniform distribution on $\theta^0_k \in [-\pi, \pi )$.
The perturbation $h(\theta)$ thus constructed follows a Gaussian distribution.
The power spectral density (PSD) decays as a power law of the wavenumber with the exponent $-(1 + 2 H)$.
This richness of the spectrum could be defined by the "magnification" parameter $\zeta$ presenting the ratio of the highest to lowest wavenumbers $\zeta = k_s / k_l$~\cite{persson2001theory}.

\begin{figure}[ht!]
    \centering
    \resizebox{1\textwidth}{!}{\input{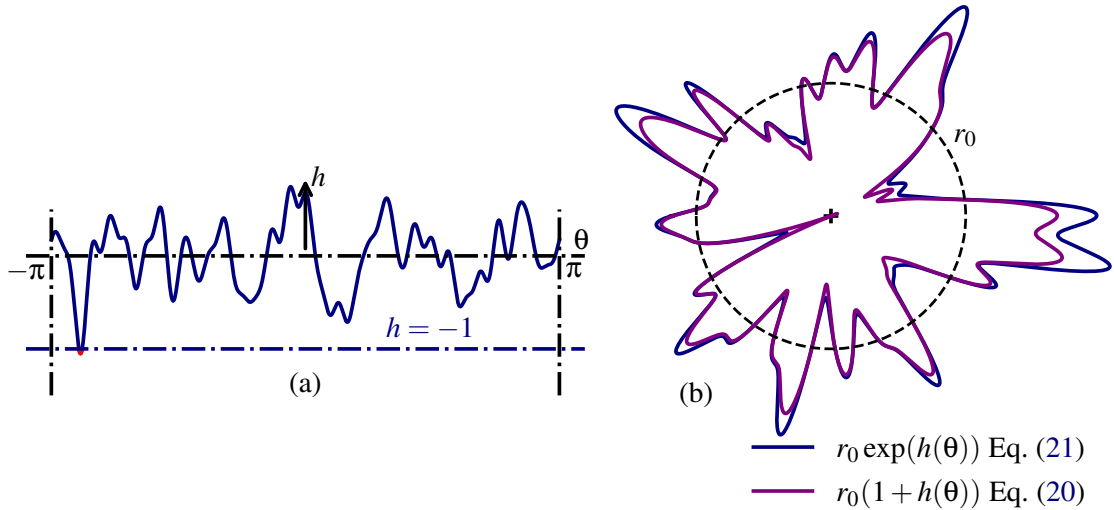} }
    \caption{\label{fig:transf_ink}Example of self-affine spot construction for $\xi = 0.25$, $k_l=4$, $k_s=32$ and $H=0.5$: (a) height perturbation $h(\theta)$ and (b) resulting spot radius for linear~\eqref{eq:rink} and exponential~\eqref{eq:rregink} transformations.
    }
\end{figure}

The radius of a contact spot in polar coordinates $r(\theta)$ can be readily defined with the perturbation $h(\theta)$ as:
\begin{equation}
    r(\theta) = r_0(1 + h(\theta)),
    \label{eq:rink}
\end{equation}
naturally $\langle r\rangle = r_0$.
Nevertheless, in this construction, even imposing $\xi < 1$ does not guarantee positivity, as the factor $1 + h(\theta)$ may become negative, consequently leading to negative values for $r(\theta)$ as well.
To overcome this problem, some smarter transformation should be used, for example:
\begin{equation}
    r(\theta) = r_0 \exp( h(\theta) ),
    \label{eq:rregink}
\end{equation}
whose two first terms of Taylor expansion is equivalent to~\eqref{eq:rink}, but the transformation~\eqref{eq:rregink} keeps the final shape well defined without self-intersections: even for $h \to -\infty$, $r \to 0$.
However, this transformation does not preserve the mean radius at $r_0$; the radius will change with $H$, $\xi$, $k_l$ and $\zeta$; this deviation will be characterized by a dimensionless quantity $\bar r = \langle r(\theta)\rangle/r_0$.
Within this formulation the parameter $\xi$ plays a similar role as in the study of flower-shaped spots: here, to the first order it presents the ratio between the amplitude of the first mode to the nominal radius.
The two transformations~\eqref{eq:rink},\eqref{eq:rregink} are illustrated in Fig.~\ref{fig:transf_ink} highlighting a situation, where for transformation~\eqref{eq:rink} the radius becomes negative. Therefore, for this study we adopt the exponential transformation~\eqref{eq:rregink}.
Several examples of complex shapes generated using the presented algorithm are shown in Fig.~\ref{fig:ex_ink} for $\xi=0.1$ and different values of $k_l$, $k_s$ and $H$ but with the same set of random phases $\theta^0_k$.
As the Hurst exponent increases, the spot naturally becomes smoother.

\begin{figure}[ht!]
    \centering
    \resizebox{1\textwidth}{!}{\input{fig_11_12_13.pgf} }
    \caption{\label{fig:ex_ink}Examples of self-affine spots and their geometrical characteristics for $\xi = 0.1$.\newline \emph{In the upper panel} (a,b,c) $k_l=2$, $k_s = \{8, 16, 32\}$, $H=0.25$: (a) $\bar{r} = 1.0017$, $\sigma = 0.057$, (b) $\bar{r} = 1.0020$, $\sigma = 0.063$, (c) $\bar{r} = 1.0023$, $\sigma = 0.067$;\newline \emph{in the middle panel} (d,e,f) $k_l=4$, $k_s = \{8, 16, 32 \}$, $H=0.25$: (d) $\bar{r} = 1.0019$, $\sigma = 0.062$, (e) $\bar{r} = 1.0029$, $\sigma = 0.075$, (f) $\bar{r} = 1.0036$, $\sigma = 0.084$;\newline \emph{in the lower panel} (g,h,i) $k_l=4$, $k_s = 128$, $H= \{0.25, 0.5, 0.75 \}$: (g) $\bar{r} = 1.0044$, $\sigma = 0.094$, (h) $\bar{r} = 1.0028$, $\sigma = 0.074$, (i) $\bar{r} = 1.0020$, $\sigma = 0.063$.
    }
\end{figure}

\subsection{Geometrical characteristics}

\begin{figure}[ht!]
    \centering
   	\resizebox{1\textwidth}{!}{\input{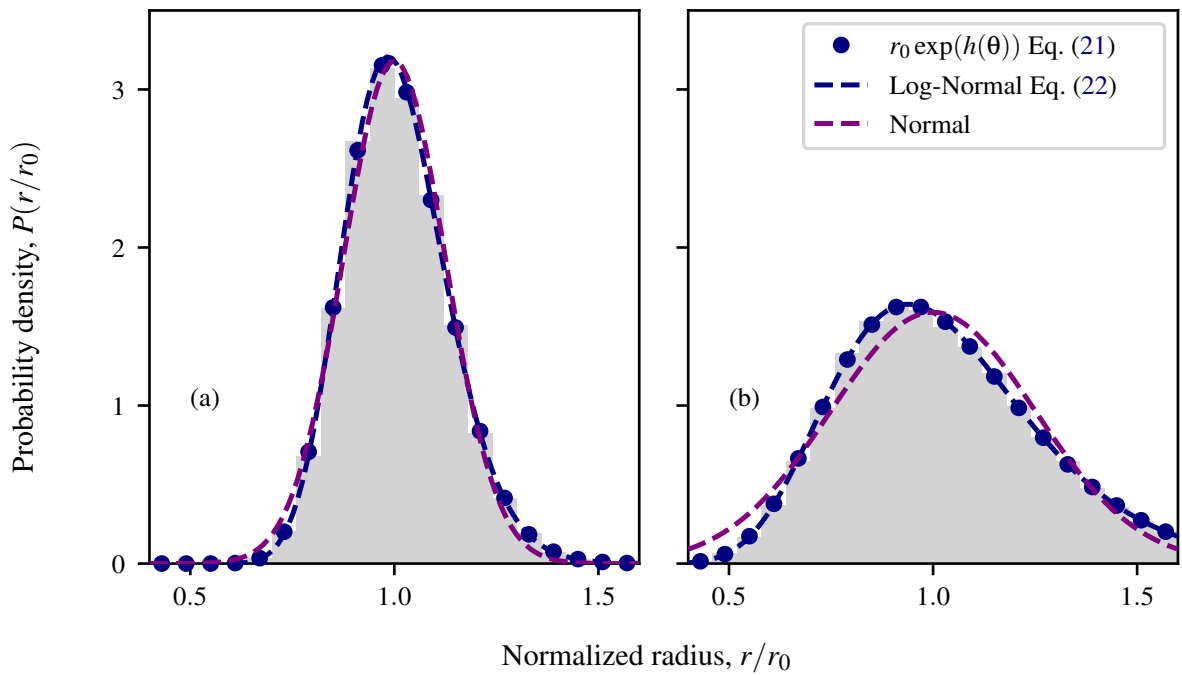}}
    \caption{\label{fig:hist_ink} Distribution of the radius for constructed self-affine spots for $H=0.25$, $k_l = 8$, $k_s = 16$ and (a) $\xi =  0.05$ and (b) $\xi = 0.1$. Least squares fit of the normal and log-normal distributions is also presented.
    }
\end{figure}

The initial height perturbation $h$ follows normal distribution with zero mean and standard deviation $\sigma_h$, but it is not preserved by the exponential transformation~\eqref{eq:rregink}.
The obtained radius follows a \textit{log-normal} distribution
\begin{equation}
    P(r) = \frac{1}{r \sigma_h \sqrt{2 \pi} } \exp{\left( -\frac{\ln^2(r)  }{2 \sigma_h^2} \right) }.
    \label{eq:pdflog}
\end{equation}
Histograms shown in Fig.~\ref{fig:hist_ink} present the probability density of the spot radius constructed for $H=0.25$, $k_l = 8$, $k_s = 16$ and two values of $\xi = \{0.05, 0.1\}$ 
and computed over 1000 generated spots, the least squares fitted normal and log-normal densities are also plotted.
For small values of $\xi$, the distribution is very close to the normal one, whereas for a higher value, it clearly follows the log-normal one.

The standard deviation $\sigma_h$ or the variance $\sigma_h^2$ of the height perturbation $h(\theta)$ can be computed taking into account the orthogonality of cosine functions:
\begin{equation}
    \sigma_h^2 = \frac{ \xi^2 }{2} \sum_{k=k_l}^{k_s} \left( \frac{k}{k_l} \right) ^{-(1 + 2 H)}
    \label{eq:spec_gauss}
\end{equation}
After the transformation, the mean radius $\langle r \rangle$ and the variance of the height distribution $\sigma_r^2$ can be found as 
\begin{subequations}
\label{eq:mu_sig_log}
\begin{align}
    \langle r \rangle  &= r_0 \exp\left( \sigma_h^2 / 2 \right), \label{eq:mu_log} \\
    \sigma_r^2 &= \left[ \exp\left(\sigma_h^2 \right) -1 \right] \exp( \sigma_h^2 ), \label{eq:sig_log}
\end{align}
\end{subequations}
The mean value of the radius is no longer equal to $r_0$, but it tends to $\langle r\rangle \to r_0$ as $\sigma_h\to0$.
Note that the variance could be expressed through the first terms of the Taylor expansion as 
$\sigma_r^2 \approx \sigma_h^2 + 1.5\sigma_h^4 + 7/6 \sigma_h^6 + O(\sigma_h^8)$, which demonstrates that for small values of $\sigma_h$, $\sigma_r \approx \sigma_h$ with a high accuracy.
The comparison of the analytical expression of the variance Eq.~\eqref{eq:sig_log} with numerically evaluated standard deviation is presented in Appendix~\ref{app:geom_selfaffine}.

The standard deviation of radius is an important geometric characteristic of the spot.
However, the standard deviation of the gradient and Laplacian of the radius as well as Nayak parameter~\cite{nayak1971random} could also have an effect on the conductivity of the spot.
These geometrical characteristics are related to spectral moments $m_0$, $m_2$ and $m_4$ as follows:
\begin{subequations}
\label{eq:m0_m2_m4}
\begin{align}
    m_0 &= \sigma^2 , \label{eq:m0} \\
    m_2 &= \langle | \nabla r |^2 \rangle = \frac{1}{2\pi}\int_0^{2 \pi} \left( \frac{1}{r(\theta)}  \frac{\partial r}{\partial \theta} \right)^2 d \theta , \label{eq:m2} \\
    m_4 &= \frac{1}{2\pi}\langle | \Delta r |^2 \rangle = \int_0^{2 \pi}  \left(\frac{1}{r^2(\theta)} \frac{\partial^2 r}{\partial\theta^2} \right)^2 d \theta , \label{eq:m4}
\end{align}
\end{subequations}
In the limit of infinitesimal perturbations $\xi \ll 1$, we can use the following analytical equations for the 2nd and 4th spectral moments (the 0th moment is nothing but the variance of radius computed in~\eqref{eq:sig_log}):
\begin{equation}
 \quad m_2 = \frac{ \xi^2}{2} \sum_{k_l}^{k_s} k^2 {\left( \frac{k}{k_l} \right) }^{-(1+2H)} , \quad m_4 = \frac{ \xi^4}{2 r_0^2} \sum_{k_l}^{k_s} k^4 {\left( \frac{k}{k_l} \right) }^{-(1+2H)} 
\label{eq:m0_m2_m4_analytical}
\end{equation}
The comparison of Eq.~\eqref{eq:m0_m2_m4_analytical} with numerically evaluated moments is presented in Appendix~\ref{app:geom_selfaffine}.
For flower-shaped spots, the spectral moments simplify to the following forms:
\begin{equation}
    m^{\text{\tiny f}}_0 = \frac{r_1^2}{2} = \frac{\xi^2 r^2_0}{2}, \quad
    m^{\text{\tiny f}}_2 = \frac{r_1^2 n^2}{2 r_0^2} = \frac{\xi^2 n^2}{2} = \frac{n'^2}{2}, \quad
    m^{\text{\tiny f}}_4 = \frac{r_1^2 n^4}{2 r_0^4} = \frac{\xi^2 n^4}{2 r_0^2 } = \frac{n'^2 n^2}{2 r_0^2 }
    \label{eq:m_0_2_4_f}
\end{equation}
Since the normalized conductivity for flower-shaped spots was shown to depend exclusively on $n' = \xi n$ (see Eq.~\eqref{eq:Q_fit}), we could suggest that for a similar normalization, the main characteristic affecting the conductivity of self-affine spots will be the standard deviation of the gradient $\sqrt{\langle | \nabla r |^2 \rangle} = \sqrt{m_2}$.
In addition, it could be shown that the area of a spot is an affine function of $m_0$: $A \approx \pi r_0^2(1+ a m_0)$, and its perimeter is an affine function of $\sqrt{m_2}$: $P \approx 2\pi r (1 + b \sqrt{ m_2} )$, where $a,b$ are positive constants.

\subsection{Normalization and result}

In contrast to the study of flower-shaped spots, this study is no longer deterministic and requires taking into account the randomness of the studied geometries.
The total flux for a given geometry $Q_i$, defined as an independent event, is assumed to have a consistent average value $\mu(Q)$ and a standard deviation $\sigma(Q)$ across the same set of spot-generative parameters.
This study aims to understand the average behavior based on these parameters.
To achieve this, we compute average values for numerous realization of spot geometries.
However, this yields only an approximate value for $\langle Q \rangle$, which depends on the number of realizations $n$.
In practice, the standard deviation of the mean value scales as $\sigma(\langle Q \rangle) = \sigma(Q) / \sqrt{n}$.
The Bienaymé-Chebyshev inequality aids in establishing a confidence interval implying a parameter $\gamma \in (0,1)$: the probability to find the mean value $\langle Q\rangle$ outside the interval $\pm \sigma(Q)/\sqrt{n\gamma}$ around its theoretical value $\mu(Q)$ is smaller than $\gamma$ whatever the true underlying distribution, i.e.
\begin{equation}
    P \left( |\mu(Q)  - \langle Q \rangle| \ge  \frac{\sigma(Q)}{ \sqrt{ n \gamma } } \right) \leq \gamma,
    \label{eq:BT}
\end{equation}
Equivalently, the probability is $(1 - \gamma)$ to find the mean value in the confidence interval $\pm \sigma(Q)/\sqrt{n\gamma}$.
When there are $11$ simulations, the interval of confidence spans approximately $\pm 3 \sigma(Q)$ with the probability of $99 \%$, effectively encapsulating the mean value $\langle Q \rangle$. 
In order to reduce the interval to one standard deviation with the same accuracy, i.e. $\sqrt{n\gamma} = 1$ for $\gamma = 0.01$ one would need to carry out $n=100$ simulations which is computationally expensive in view of the number of parameters to be studied.
For $n=11$ the probability to find the mean value in one standard deviation interval, i.e. $\sqrt{11\gamma} = 1$ is higher than $\approx$9\%.
While the Bienaymé-Chebyshev inequality provides a rigorous lower bound, the actual accuracy of our results can overpass this conservative limit.
To balance the computational efforts and the accuracy, the number of BEM simulation results per combination of parameters was set to $n=11$, and this dataset was utilized to estimate the mean value and the confidence interval obtained from the measurement of the standard deviation. As seen in practice for this problem the standard deviation is much smaller than the approximate mean value, therefore the confidence interval is very narrow and $n=11$ seems to be a good compromise.

\begin{figure}[ht!]
    \centering
    \includegraphics[width=1\textwidth]{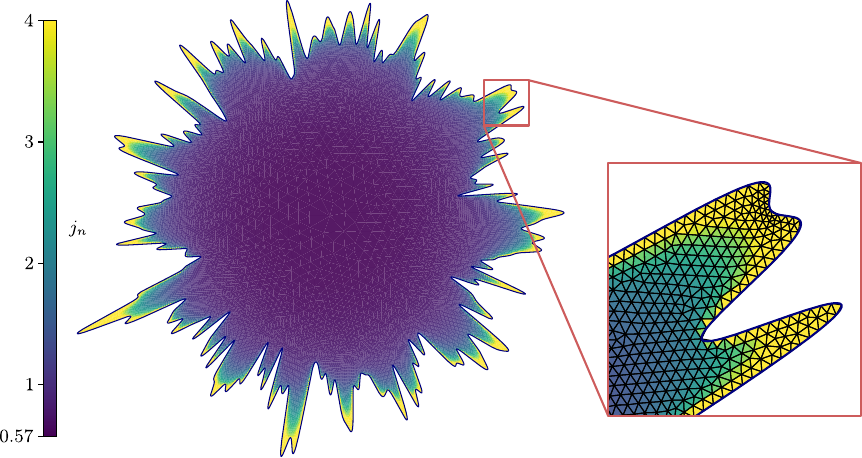}
	\caption{\label{fig:j_ink1} Example of simulation results representing the flux distribution for a self-affine spot with $\xi=0.05$, $k_l = 8$, $k_s = 128$ and $H=0.25$.}
\end{figure}

An example of flux distribution obtained by fast-BEM is displayed in Fig.~\ref{fig:j_ink1} for $\xi=0.05$, $k_l = 8$, $k_s = 128$ and $H=0.25$.
The normal flux remains singular at the edge but less pronounced at troughs than at crests as shown previously for the flower-shaped contact spot.
Compared to the latter, it appears more difficult to construct a good mesh for self-affine spots efficiently (fine mesh near the border and coarse far from border).
The mesh size was prescribed as a function of the edge curvature and as a function of the shortest distance to the border.
See zoom in Fig.~\ref{fig:j_ink1}, the finest used mesh reach $N_e = 34\,340$ elements.
As previously, to employ the Richardson extrapolation, two meshes of different density were used to obtain accurate results.

For the global flux analysis, the following geometrical characteristics describe sufficiently well the geometry
$\mathcal A =\{\langle r\rangle,\sqrt{m_0},\sqrt{m_2},\sqrt{m_4},H\}$: (1) the mean radius $\langle r\rangle$, 
the standard deviation of (2) radius $\sqrt{m_0}$, 
(3) of its gradient $\sqrt{m_2}$, 
(4) of its Laplacian $\sqrt{m_4}$ and (5) the Hurst exponent.
The initial set of independent generative parameters is $\mathcal I = \{r_0,\xi,k_l,k_s,H\}$ and contains (1') radius $r_0$, (2') amplitude of perturbation $\xi$, (3') lower $k_l$ and (4') upper $k_s$ cutoffs, and (5) the Hurst exponent $H$.
The set $\mathcal A$ is easy to measure for any spot, and it allows to study the effect of individual parameters on the total flux; the generative set $\mathcal I$ is complete and dimensionless, but suffers from the fact that it could not be easily derived for an arbitrary geometry.
A sensitivity analysis will be carried out on both sets.

The first task is to normalize the total flux $Q$ produced by self-affine spots.
There are two main options: it can be normalized by the flux of an equivalent circular contact spot (a) of the same mean radius $Q_{\circ} = 4 k U_0 \langle r \rangle$ or (b) of the same area.
The first option was selected since the mean radius also enters the set of parameters, moreover $Q_\circ$ is equivalent to the definition of $Q_{\min}$ used for the flower-shaped contact spot.
The normalized total flux is thus defined by
\begin{equation}
    Q' = \frac{Q}{ Q_{\circ} } = \frac{Q}{4kU_0\langle r\rangle}
    \label{eq:Q_ink_norm}
\end{equation}
Therefore, since the problem does not have an internal length, we can exclude the mean radius $\langle r\rangle$ from the set of parameters defining the total flux $Q'$ and consider two sets of dimensionless parameters:
\begin{itemize}
 \item Geometrical set of parameters $\mathcal A': = \{\sqrt{m_0}/\langle r\rangle, m_2, \langle r\rangle\sqrt{m_4},H\}$
 \item Generative set of parameters $\mathcal I' = \{\xi,k_l,k_s,H\}$
\end{itemize}

By analogy with the maximal total flux $Q_{\text{\tiny up}}$ defined for flower-shaped spots, we could define an equivalent upper limit for self-affine spots.
It is not a good idea to define it as the maximal radius of the self-affine spot as it could tend to infinity for very high number of modes.
However, the variance of the flower-shaped spot $m^{\text{\tiny f}}_0$, Eq.~\eqref{eq:m_0_2_4_f}, provides us with a hint getting back the half-petal length $r_1$ in another way, 
i.e. using this equation we can express it as $r_1 = (2m_0^{\text{\tiny f}})^{1/2}$; therefore, the limit characteristic radius could be expressed as $\langle r\rangle + \sqrt{2m_0}$.
Thus, the difference between two flows used for normalization takes the following form:
\[
Q_{\text{\tiny up}} - Q_{\circ} = 4kU_0\sqrt{2m_0}
\]
Following the same renormalization between zero and one as was used for the flower-shaped spots, we could define the renormalized flux as:
\begin{equation}
 Q'' = \frac{Q - Q_{\circ}}{ Q_{\text{\tiny up}} - Q_{\circ} } = \frac{Q}{4kU_0\sqrt{2m_0}} - \frac{\langle r\rangle}{\sqrt{2m_0}}
 \label{eq:Q_renorm}
\end{equation}

\begin{figure}[ht!]
    \centering
    \includegraphics[width=1\linewidth]{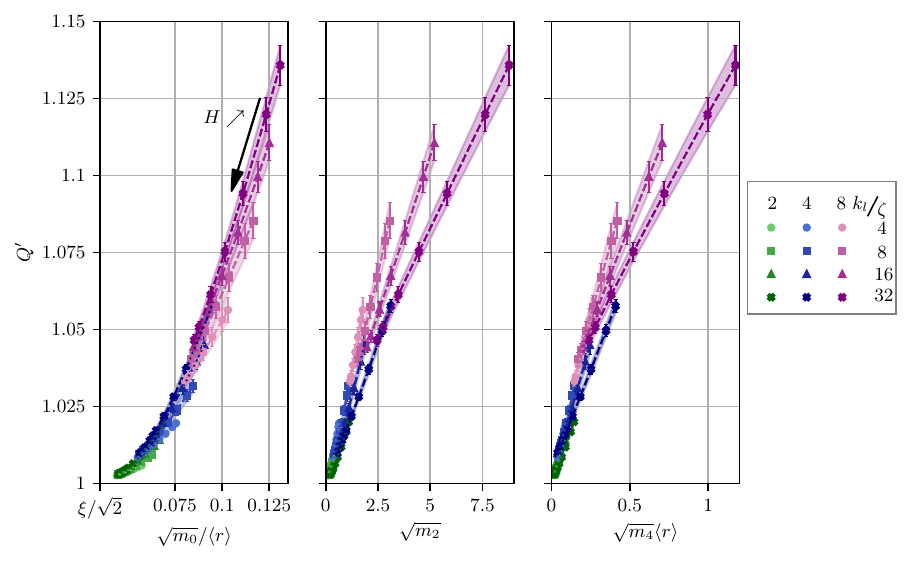}
    \caption{ \label{fig:Q_ink} Normalized flux~\eqref{eq:Q_ink_norm} of self-affine spots: for the following generative parameters $\mathcal I'$: $\xi=0.05$, $k_l = \{2, 4, 8 \}$, $\zeta = \{4, 8, 16, 32\}$ and $H = \{0.25, 0.3, 0.4, 0.5, 0.6, 0.7, 0.75\}$, note that $k_s = \zeta k_l$.
The mean normalized flux is plotted with a marker and shaded around according to the confidence interval half-width of $\pm 1.35 \sigma$ defined for $\gamma=0.05$ (namely with a rate of confidence of $95 \%$).}
\end{figure}

The total flux normalized according to Eq.~\eqref{eq:Q_ink_norm} is presented in Fig.\ref{fig:Q_ink} for all simulated self-affine spots.
These spots are constructed by changing the lower cutoff $k_l = \{2, 4, 8 \}$ and for four values of the upper cutoff $k_s = \zeta k_l$ with the magnification $\zeta = \{4, 8, 16, 32\}$.
The Hurst parameter $H$ takes the values $H = \{0.25, 0.3, 0.4, 0.5, 0.6, 0.7, 0.75\} $).
The colors are used to distinguish the 3 sets of the results according to different $k_l$.
In each color set, the results are distinguished by their marker style according to values of magnification $\zeta$, moreover, the higher the $\zeta$, the darker the color.
Along every result-curve the Hurst exponent $H$ changes as shown by the arrow: the smaller the $H$, the higher the flux.
The curves are entwined together, but they seem to follow the same trend.
Plotting the data with respect to geometrical parameters $\mathcal A'$ offers a better representation than the use of the generative set of parameters $\mathcal I'$.
The variation in slope seems to be controlled by parameter $\zeta = k_s/k_l$: increasing $\zeta$ increases the average slope with respect to $\sqrt{m_2}/\langle r\rangle$ and decreases the average slope with respect to $\sqrt{m_2}$ and $\langle r\rangle \sqrt{m_4}$.

\begin{figure}[ht!]
    \centering
    \includegraphics[width=1\linewidth]{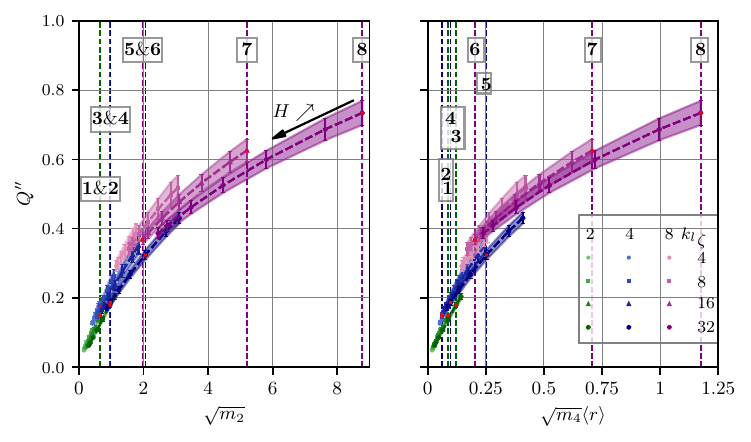}
    \caption{\label{fig:Q_ink2}
Renormalized flux, Eq.~\eqref{eq:Q_renorm} of self-affine spots: for $\xi=0.05$, $k_l = \{2, 4, 8 \}$, $\zeta = \{4, 8, 16, 32\}$, $H = \{0.25, 0.3, 0.4, 0.5, 0.6, 0.7, 0.75\}$.
The mean normalized flux is plotted with a marker and shaded around according to the confidence interval half-width of $\pm 1.35 \sigma$ and of $95 \%$ of accuracy; vertical lines and numbers correspond to spots shown in Fig.~\ref{fig:ink_spots}.}    
\end{figure}

The results of the renormalized flux Eq.\eqref{eq:Q_renorm} with respect to $\sqrt{m_2}$ and $\langle r\rangle\sqrt{m_4}$ are presented 
in Fig.~\ref{fig:Q_ink2}.
However, the role of the fourth moment in the form $\langle r\rangle\sqrt{m_4}$ seems to be strongly correlated with $\sqrt{m_2}$ and does not bring much additional information.
The lack of simple dependency of the normalized flux with respect to geometrical characteristics pushes us to suggest an alternative normalization.
The shown results seem to depend strongly on  $k_l$ parameter.
By exploring a wider spreading of $\sqrt{m_2}$ for different $k_l$, we hope to easier identify their influence, which is the objective of the following sections.

\begin{figure}[htb!]
    \centering
    \includegraphics[width=0.49\textwidth]{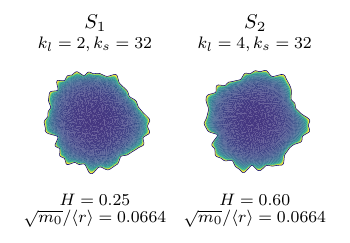}
    \hspace{.0\textwidth}%
    \includegraphics[width=0.49\textwidth]{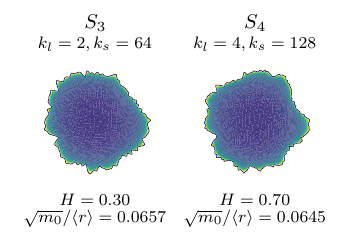}\\[1em]
    \includegraphics[width=0.49\textwidth]{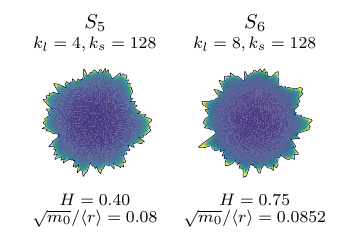}
    \hspace{.0\textwidth}%
    \includegraphics[width=0.49\textwidth]{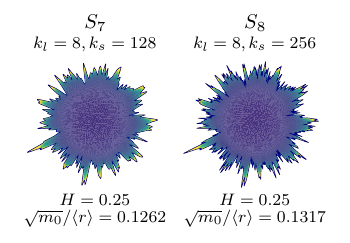}
    \caption{\label{fig:ink_spots} Examples of self-affine spots with corresponding generative characteristics; the corresponding data points are highlighted in Fig.~\ref{fig:Q_ink2}, the resulting flux and geometrical characteristics can be found in Table~\ref{tab:spots18_params}.}
\end{figure}

To provide a visual geometrical interpretation in the flux variation (see Fig.~\ref{fig:ink_spots}) related to geometrical characteristics of self-affine spots, 
we present particular shapes along with the values of corresponding geometrical characteristics and of the total flux in Table~\ref{tab:spots18_params}, 
the location of these particular spots is also highlighted in Fig.~\ref{fig:Q_ink2}.
The pairs $ \{ S_1,S_2 \}$, $\{ S_3,S_4 \}$ and $\{ S_5,S_6 \}$ have close values of $\sqrt{m_2}$.
The difference between the flux of spots $S_5$ and $S_6$ highlights the fact that the flux does not depend only on the second moment.
Nevertheless, this could be seen as a second order effect compared to that of $\sqrt{m_2}$.
Spots $S_7$ and $S_8$ are among the "roughest" spots and possess the highest flux.

\begin{table}[htb!]  
\begin{center}
\begin{tabular}{ cllllllll }
  &  & \multicolumn{3}{c}{ Parameters $ \mathcal A' $} & \multicolumn{4}{c}{Parameters $\mathcal I'$} \\
 Spot \# & $Q''$ & $\sqrt{m_0} / \langle r \rangle$ & $\sqrt{m_2}$ & $\sqrt{m_4} \langle r \rangle $ & $\xi$ & $k_l$ & $k_s$ & $H$ \\
 \hline
 $S_1$ & $0.151$ & $0.0664$ & $0.658$ & $14$ & $0.05$ & $2$ & $32$ & $0.25$ \\
 $S_2$ & $0.172$ & $0.0664$ & $0.666$ & $12.7$ & $0.05$ & $4$ & $32$ & $0.6$ \\
 $S_3$ & $0.173$ & $0.0657$ & $0.958$ & $39.8$ & $0.05$ & $2$ & $64$ & $0.3$ \\
 $S_4$ & $0.184$ & $0.0645$ & $0.972$ & $63.8$ & $0.05$ & $4$ & $128$ & $0.7$ \\
 $S_5$ & $0.317$ & $0.0800$ & $2.06$ & $163$ & $0.05$ & $4$ & $128$ & $0.4$ \\
 $S_6$ & $0.380$ & $0.0852$ & $1.97$ & $130$ & $0.05$ & $8$ & $128$ & $0.75$ \\
 $S_7$ & $0.644$ & $0.126$ & $5.12$ & $446$ & $0.05$ & $8$ & $128$ & $0.25$ \\
 $S_8$ & $0.749$ & $0.182$ & $8.77$ & $1493$ & $0.05$ & $8$ & $256$ & $0.25$ \\
\end{tabular}
\end{center}
\caption{\label{tab:spots18_params}Parameters for spots shown in Fig.~\ref{fig:ink_spots} and the resulting total flux. }
\end{table}

\subsection{Results with renormalized standard deviation}

As presented in Figs.~\ref{fig:Q_ink} and \ref{fig:Q_ink2}, the results are clustered with respect to $k_l$.
To have more control on geometrical characteristics, we renormalize the generative function $h(\theta)$ in order to prescribe its dimensionless standard deviation $\sigma_h = \sqrt{m_{0,h}}$ (see Eq.~\eqref{eq:spec_gauss}):
\begin{equation}
\label{eq:h_m0}
    h( \theta ) =  \sqrt{ \frac{2m_{0,h}}{s} } \sum_{k=k_l}^{k_s} \xi_k \cos{(k \theta + \theta^0_k )}, \quad s = \sum_{k=k_l}^{k_s} \xi_k^2 
\end{equation}
with $\xi_k$ defined by Eq.~\eqref{eq:r_k}.
The exponential transformation from $h(\theta)$ to radius $r(\theta)$ ~\eqref{eq:rregink} remains intact.
Then, for the normalized generative parameters we have: $\mathcal I'' = \{\sqrt{m_{0,h}}, k_l, k_s, H\}$ and for the geometrical ones we still have $\mathcal A'$.
The main goal for such a choice is to decorrelate  $\sqrt{m_0}/\langle r\rangle$ and $\sqrt{m_2}$
and thus to level down $m_0$ for $k_l=8$, and to level it up for $k_l=2$.
Note also that $\sqrt{m_0}/\langle r\rangle \approx \sqrt{m_{0,h}}$ for small values of the latter, see~\eqref{eq:sig_log} and its Taylor expansion.

\begin{figure}[htb!]
    \centering
    \includegraphics[width=1\textwidth]{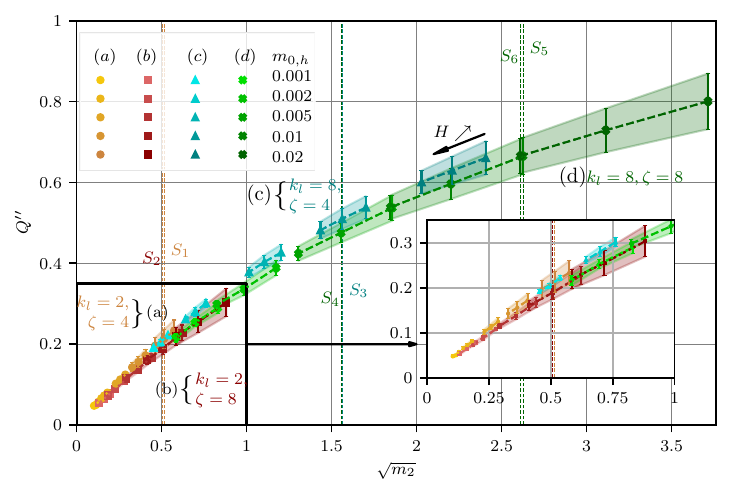}
    \caption{\label{fig:Q_ink2_m0} Results for the renormalized total flux Eq.~\eqref{eq:Q_renorm} of self-affine spots with respect to standard deviation of the radius gradient $\sqrt{m_2}$ obtained with controlled standard deviation $\sqrt{m_{0,h}} \approx \sqrt{m_0}/\langle r\rangle$.
    }
\end{figure}

The results of this set of simulations for the renormalized flux~\eqref{eq:Q_renorm} are presented in Fig.~\ref{fig:Q_ink2_m0} with respect to the standard deviation of the radius gradient $\sqrt{m_2}$;  lower cutoffs $k_l=2$ (orange circles and red squares) and $k_l=8$ (green crosses and cyan triangles) were used.
In addition, different colors correspond to different magnifications: $\zeta=4$ for red squares and cyan triangles, $\zeta=8$ for orange circles and green crosses.
The Hurst exponent takes three values $H=\{0.25, 0.50, 0.75\}$.
For the same value of $\sqrt{m_{0,h}} \approx \sqrt{m_0}/\langle r\rangle$, thanks to the variation of the Hurst exponent, the value of $\sqrt{m_2}$ varies within a certain interval, such data points are connected by a line.
Every set of such lines (of the same shade) align along their master curve.
Such results demonstrate that even though the standard deviation $\sqrt{m_0}/\langle r\rangle$ controls the thermal flux to a large extent, the standard deviation of the gradient $\sqrt{m_2}$ also influences the result.
This conclusion is possible since $\sqrt{m_2}$ does not enter the flux normalization~\eqref{eq:Q_renorm}.
The flux increases with respect to both $\sqrt{m_0}/\langle r\rangle$ and $\sqrt{m_2}$ as well as with respect to $\langle r\rangle\sqrt{m_4}$ because of the strong correlation of the latter with the second moment.
For an equivalent $\sqrt{m_2}$, the flux is higher for spots with a lower magnification $\zeta$.

\begin{figure}[htb!]
    \centering
    \includegraphics[width=0.49\textwidth]{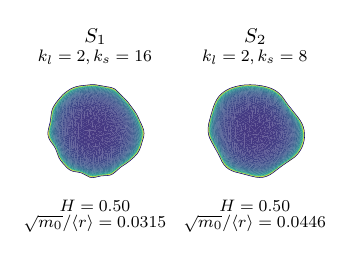}
    \hspace{.0\textwidth}%
    \includegraphics[width=0.49\textwidth]{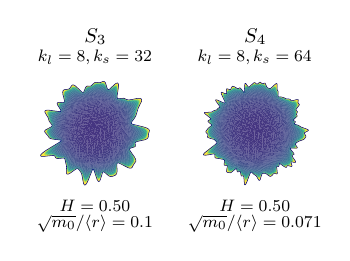} \\[1em]
    \includegraphics[width=0.49\textwidth]{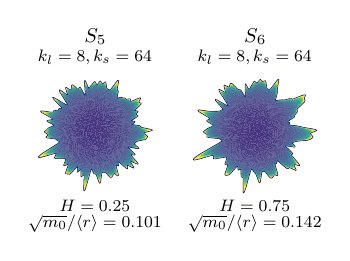}
    \caption{\label{fig:spot12_34_56_m0} Comparison of pairs of self-affine spots for the same $\sqrt{m_2}$ and for different values of $\sqrt{m_{0}}/\langle r\rangle $.
To simplify the reading of the parameters, they are equivalently displayed in Table~\ref{tab:spots_params}.
    }
\end{figure}

To provide a geometrical meaning to these results, coupled pairs of self-affine spots $\{S_1,S_2\},\{S_3,S_4\},\{S_5,S_6\}$ with different spectral content but similar value of $m_2$ are displayed in Fig.~\ref{fig:spot12_34_56_m0} and are highlighted in Fig.~\ref{fig:Q_ink2_m0} and in Table~\ref{tab:spots_params}.
Remarkably, the three foremost right lines (d) in the figure seem to continue each other.
The spots $\{S_5,S_6\}$ well illustrate this link: they in fact have the same number of modes, but different $H$ and $m_{0,h}$.

\begin{table}[htb!]   
\begin{center}
\begin{tabular}{ cllllllll }
  &  & \multicolumn{3}{c}{ Parameters $ \mathcal A' $} & \multicolumn{4}{c}{Parameters $\mathcal I''$} \\
 Spot \# & $Q''$ & $\sqrt{m_0} / \langle r \rangle$ & $\sqrt{m_2}$ & $\sqrt{m_4} \langle r \rangle $ & $\sqrt{m_{0,h}}$ & $k_l$ & $k_s$ & $H$ \\
 \hline
 $S_1$ & $0.189$ & $0.100$ & $0.507$ & $5.10$ &  $0.100$ & $2$ & $16$ & $0.5$ \\
 $S_2$ & $0.213$ & $0.143$ & $0.515$ & $2.84$ & $0.141$ & $2$ & $8$ & $0.5$ \\
 $S_3$ & $0.510$ & $0.0995$ & $1.56$ & $33.6$ & $0.100$ & $8$ & $32$ & $0.5$ \\
 $S_4$ & $0.476$ & $0.0705$ & $1.56$ & $61.9$ & $0.005$ & $8$ & $64$ & $0.5$ \\
 $S_5$ & $0.665$ & $0.0997$ & $2.62$ & $114$ & $0.100$ & $8$ & $64$ & $0.25$ \\
 $S_6$ & $0.666$ & $0.141$ & $2.61$ & $94.1$ & $0.141$ & $8$ & $64$ & $0.75$ \\
\end{tabular}
\end{center}
\caption{\label{tab:spots_params}Parameters for spots shown in Fig.~\ref{fig:spot12_34_56_m0} and the resulting total flux.
The spots are ordered according to the increasing value of $\sqrt{m_0}$.
}
\end{table}

Finally, we would like to point out that (1) the influence of $\sqrt{m_0}$ is ultimately handled by normalization;
 (2) the results are rather well clustered along a simple trend line in terms of $m_2$,
 (3) but clearly there is a dependence on $\zeta$. At the same time, the shape of the trend is very similar to what was observed for multi-petal spots and we recall that $\sqrt{m_2}$ is analogous parameter to $n'$ used there. Qualitatively the slope of the normalized total flux with respect to $\sqrt{m_2}$ decreases suggesting an ultimate saturation as in multi-petal shapes (see Figs.~\ref{fig:Q_flower} and \ref{fig:Qn_star_gear}). 
 For the extra generative parameter $\zeta \in\mathcal I''$ to which some dependence is observed, it should be expressed through spectral moments which could be easily measured for arbitrary shape, it will be handled in the following subsection.

\subsection{Conductivity model}

This study aims to quantify the flux transmitted through a spot of complex shape. 
While the numerical results encompass a broad parametric space, they are not readily comprehensible in their full scope and pose challenges for generalization.
We thus make an attempt to construct a general phenomenological model relying on geometrical characteristics and inspired from the model used for flower-shaped spots.

\subsubsection{Covariance matrix}

The first simple step is the construction of a covariance $C_{ij}$ matrix of the normalized flux and all available normalized parameters $\tilde x_i$:
\begin{equation}
 C_{ij} = \langle \tilde x_i \tilde x_j \rangle, \quad \tilde x_i = \frac{x_i - \langle x_i\rangle}{\sigma(x_i)},
 \label{eq:cov_matrix}
\end{equation}
where, as previously, $\langle x_i\rangle$ denotes the average value, and $\sigma(x_i)$ denotes its standard deviation.
The covariance matrix constructed based on all available simulation data is provided in Table~\ref{tab:covariance}.
There is a strong correlation between $\tilde Q'$, $\tilde Q''$ and parameters $\tilde k_l$, $\tilde{\sqrt{m_0}/\langle r\rangle}$, $\tilde{\sqrt{m_2}}$ and $\tilde{\sqrt{m_4}\langle r\rangle}$.
However, because of the strong correlation, the effect of the moment $m_4$ is hard to isolate from the effect of $m_2$.
Very small correlation of the flux is found with $\tilde\xi$, $\tilde H$ and $\tilde\alpha$; slightly more correlation exists with $\tilde\zeta$. According to the covariance matrix, the Hurst exponent seems to be negligible, which is surprising in the light of our previous results. In conclusion, we could confirm that the covariance matrix and eventual Principal Component Analysis, which access only first order correlations, present too coarse tools to determine subtle non-linear correlations.
Finally, since generative parameters $\mathcal I$ are strongly linked to the method of spot generation, in constructing our model we will focus exclusively on geometrical parameters $\mathcal A$ which could be measured for an arbitrary shape.

\begin{table}
\begin{tabular}{ccccccccccc}
 & $\tilde Q'$ & $\tilde Q''$ & $\tilde \xi$ & $\tilde k_l$ & $\tilde \zeta$ & $\tilde H$
 & $\tilde{\sqrt{m_0}/\langle r\rangle}$    & $\tilde{\sqrt{m_2}}$ & $\tilde{\sqrt{m_4} \langle r\rangle}$ & $\tilde\alpha$\\
  $\tilde Q'$                               & 1 & 0.9 & 0.1 & 0.7 & 0.3 & -0.2 & 0.7 & 0.9 & 0.7 & 0.06\\
  $\tilde Q''$                              &-& 1 & -0.04 & 0.8 & 0.4 & -0.2 & 0.7 & 0.9 & 0.7 & 0.1 \\
  $\tilde \xi$                              &-&-& 1 & -0.2 & -0.2 & 0.1 & 0.7 & -0.06 & -0.07 & -0.08 \\
  $\tilde k_l$                              &-&-&-& 1 & 0.06 & 0.0 & 0.4 & 0.7 & 0.4 & -0.03 \\
  $\tilde \zeta$                            &-&-&-&-& 1 & 0.0 & 0.06 & 0.5 & 0.5 & 0.9\\
  $\tilde H$                                &-&-&-&-&-& 1 & -0.2 & -0.2 & 0.3 & 0.2 \\
  $\tilde{\sqrt{m_0}/\langle r\rangle}$     &-&-&-&-&-&-& 1 & 0.6 & 0.4 & -0.05\\
  $\tilde{\sqrt{m_2}}$                      &-&-&-&-&-&-&-& 1 & 0.9 & 0.2\\
  $\tilde{\sqrt{m_4} \langle r\rangle}$     &-&-&-&-&-&-&-&-& 1 & 0.3\\
  $\tilde\alpha$                            &-&-&-&-&-&-&-&-&-& 1\\
\end{tabular}
\caption{\label{tab:covariance}Covariance matrix of normalized flux and all normalized parameters according to Eq.~\eqref{eq:cov_matrix}.}
\end{table}

\subsubsection{Phenomenological model}

Analyzing the obtained results, we noticed a weak logarithmic dependence of the total normalized flux on the magnification parameter $\zeta = k_s/k_l$. A relatively simple phenomenological model including this parameter could be constructed, but since this parameter is generative, it is of no help for a general case. Nevertheless, it is clear that the magnification $\zeta$ is intimately related to another geometrical parameter, known as Nayak parameter~\cite{nayak1971random,yastrebov2017role} $\alpha = m_0m_4/m_2^2$ (see covariance matrix in Table~\ref{tab:covariance}). Remark that from Eqs.~\eqref{eq:m_0_2_4_f} for the flower-shaped spot, the Nayak parameter is simply $1$ so, consistently it does not enter the phenomenological equation for the conductivity of such simple forms Eq.~\eqref{eq:Q_fit}.
A rigorous link between the generative parameter $\zeta$ and the geometrical parameter $\alpha$ could be provided, see Appendix~\ref{app:geom_selfaffine}. The concrete form of this link was not used, but we could notice that another geometrical characteristic, namely the Hurst exponent $H$ is involved, regardless the results of the covariance analysis. So, the ultimate set of geometrical dimensionless parameters is chosen to be:
\begin{equation}
    \mathcal A^f = \left\{\sqrt{m_0}/\langle r\rangle, \sqrt{m_2}, H, \alpha\right\} \equiv \left\{\sigma/\langle r\rangle, \sqrt{\langle (\nabla r)^2\rangle}, H, \alpha\right\}
\end{equation}
The fourth moment $m_4$ does not enter explicitly in the set of parameters, only through the Nayak parameter $\alpha$ similarly to models of rough contact~\cite{greenwood2006simplified,carbone2008asperity,yastrebov2017role}.
Finally, we suggest the following form for the phenomenological model:
\begin{equation}
    Q'' = a \left[ 1 - \frac{1}{b \sqrt{m_2} + 1} \right] \left( 1 + c H \right) \left\{ 1 + \frac{ d }{ e \alpha^f + 1 } \right\}
    \label{eq:q_selfaffine_model}
\end{equation}
with the core term in square brackets which is equivalent to the phenomenological law obtained for multi-petal spots, see Eq.~\eqref{eq:Q_fit}.
In addition, the effects of $H$ and $\alpha$ enter the equation through the product of linear and non-linear functions, respectively (normal and curly brackets).
The former is the increasing function of $H$ and the latter is a decreasing function of $\alpha$.
Both terms provide a slight factor adjustment: in the interval $(1,1+c)$ for $H \in (0,1)$, and in the interval $(1+d/(1+e),1)$ for $\alpha\in(1,\infty)$.
Due to a weak dependence on the Nayak parameter, we made an attempt to integrate it through a logarithmic dependence, like in~\cite{yastrebov2017role}, 
but the constructed model could not fulfil the physical consistency, i.e. to always ensure positive normalized flux $Q''$ which increases monotonically for increasing $\alpha$ and $\zeta$ (see Appendix~\ref{app:dQ}).
This physical consistency could be formulated as an inequality for the exponent parameter $f$:
$$ f \ge \frac{1-H}{2H}.$$
The issue with this bound is that it diverges for $H \to 0$. Therefore, we deliberately fixed the minimal value of the Hurst exponent that we took into consideration $H\ge0.25$, providing the following condition for the exponent $f\ge 1.5$. A further study should be carried out to formulate a physically consistent phenomenological model for the flux for spots with lower values of the Hurst exponent.

\begin{figure}[ht!]
    \centering
    \includegraphics[width=1\textwidth]{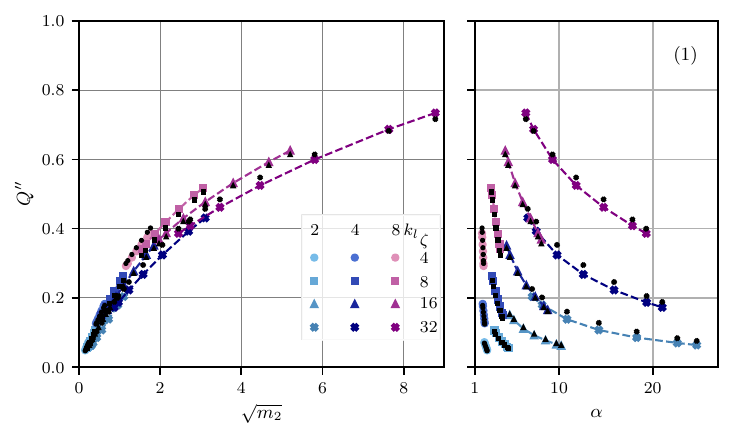}\\
    \includegraphics[width=1\textwidth]{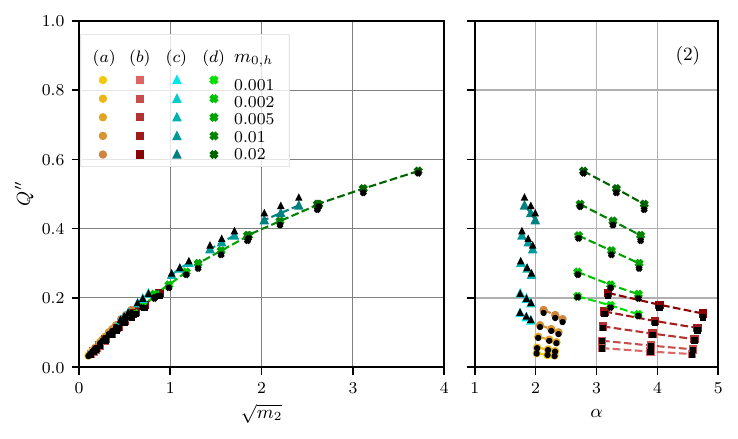}
    \caption{\label{fig:Q_ink2_inter} Simulation results for the normalized flux through self-affine spots (color markers and interpolation lines) and phenomenological prediction (smaller black markers of the same type).
Upper row: initial set of data $(1)$, lower raw: data with controlled standard deviation $(2)$.
The letters in the second series refer to those defined Fig.~\ref{fig:Q_ink2_m0}.}
\end{figure}

Combining Eqs.~\eqref{eq:q_selfaffine_model} and \eqref{eq:Q_renorm}, the final equation for the flux is obtained as:
\begin{equation}
	Q = Q_{\circ} \left( 1 + a \frac{\sqrt{2 m_0}}{\langle r\rangle} \left[ 1 - \frac{1}{ b \sqrt{m_2} + 1} \right] \left( 1 + c H \right) \left\{ 1 + \frac{d}{e \alpha^f + 1} \right\} \right) \label{eq:Q_fit_int}
\end{equation}
The coefficients are found by the least square fit of all simulation results, see Table~\ref{tab:tab_fit}.
Results of the fitting law are shown in Fig.~\ref{fig:Q_ink2_inter} separately for two sets of simulation data: in Fig.~\ref{fig:Q_ink2_inter}(a) for the set of contact spots parametrized by Eq.~\eqref{eq:h_ink}, and in Fig.~\ref{fig:Q_ink2_inter}(b) for those defined by Eq.~\eqref{eq:h_m0}.
A relative error could be defined as:
\begin{equation}
    E = \frac{1}{N} \sum\limits_{i=1}^N \frac{|Q_i - Q_i^{\text{\tiny fit}}|}{Q_i},
\end{equation}
and for the fit coefficients the error reduces to $E=4.3$ \%.
The Pearson's correlation factor for the set of identified parameters is equal to $\rho = 0.9976$.

\begin{table}[htb!]
\begin{center}
\begin{tabular}{ lllllll }
    \hline
Parameter & $a$ & $b$ & $c$ & $d$ & $e$ & $f$ \\
Value & $0.968$ & $0.255$ & $0.0867$ & $4.38$ & $5.49$ & $1.50$ \\
\hline
\end{tabular}
\end{center}
\caption{ \label{tab:tab_fit} Parameters of the phenomenological model~\eqref{eq:q_selfaffine_model},\eqref{eq:Q_fit_int} optimized through least square fit and resulting in relative error $E=4.3$ \% and Pearson's correlation factor $\rho = 0.9976$.}
\end{table}

The high concentration of data for low $\sqrt{m_2}$ might have introduced biases during the fitting process.
Nonetheless, the obtained model captures well all the trends observed in our simulation results, notably, it represents well the flux of the roughest contact spots with highest values of $\sqrt{m_2}$ and $\alpha$.
In summary, the obtained model could be seen as a generalization of the initial model formulated for multi-petal shapes (flower-, star- and gear-like). The ultimate model integrates the combined effects not only of standard deviation $\sigma = \sqrt{\langle (r-\langle r\rangle)^2\rangle} = \sqrt{m_0}$ and 
$\sqrt{\langle (\nabla r)^2\rangle} = \sqrt{m_2}$ but also of more subtle shape parameters such as the Nayak parameter $\alpha$ and the Hurst exponent $H$, which are related to bandwidth length and fractal dimension, respectively.

As a by product, the form of the phenomenological model~\eqref{eq:Q_fit_int} permits us to access the fractal limit of the self-affine spots, when the magnification
$\zeta = k_s/k_l \to \infty$, then a very simple form for the limit flux could be obtained, depending only on the standard deviation of the spot and its Hurst exponent:
\begin{equation}
  \lim\limits_{\zeta\to\infty}(Q) = Q_{\circ} \left( 1 + a \frac{\sqrt{2 m_0}}{\langle r\rangle}\left( 1 + c H \right)\right) = 4kU_0\left(\langle r\rangle +  a \sqrt{2 m_0} \left( 1 + c H \right)\right).
\end{equation}
As a first order approximation, one could use the following value $4kU_0\left(\langle r\rangle +  \sqrt{2 m_0}\right)$ 
which remains relatively accurate due to the factor $a(1+cH)$ having minimal variation, remaining within the range $(0.968,1.052)$. 
In general, this fractal limit remains speculative and could be seen as our conjecture as for the case of multi-petal shapes.

\section{Conclusion\label{sec:conclusion}}

In establishing the bounds on the conductivity of rough contacts~\cite{barber2003bounds}, Barber argued that "its greatest potential probably lies in establishing the maximum effect of neglected microscales of roughness in a solution of the contact problem for bodies with multiscale or fractal roughness."
In our contribution, we focus on these "microscales" and make an attempt to assess their quantitative effect on the conductivity.
If we repeat after Samuel Karlin that "the purpose of models is not to fit the data but to sharpen the questions",
this study indeed permitted to sharpen few of them.

\subsection{Flower-shaped spots and other simple forms}

For simple multi-petal shapes: flower-, star- and gear-like conductive spots we could obtain the following results.
In the limit of the infinite number of petals, rays, and teeth, the conductivity seems to converge to different finite limits.
The bigger the area, the higher its limit, therefore the gear-like shapes have the highest and star-like shapes the lowest conductivity.
We determined these limits by an extrapolation of a constructed phenomenological model, and these results should be interpreted as a first guess.
Hence, the \textit{first question} is whether a conductivity of such spots could be determined analytically in the limit of infinite number of petals, rays or teeth?
Expectedly, these limits are bounded between the conductivity of a circle with the average radius $r_0$ and a circle with the radius equal to the maximal extent of these spots $r_0(1+\xi)$.
On the other hand, in this limit, the boundary of the conducting spot could be seen as fuzzy, with the same geometrical bounds but different "fuzziness" types, which surprisingly significantly affects the limit.
The physical and mathematical limits could be different here because of radiative and eventually convective heat exchanges or because of tunneling effects.
The physical conductivity should probably hit the upper limit $Q_{\text{\tiny up}}$ defined by the conductivity of a circular spot of radius $r_0(1+\xi)$.

\subsection{Conductivity of self-affine spots}

In terms of conductivity of self-affine random spots, based on numerous simulation results and being inspired by the phenomenological model constructed for a flower-shaped spot, we suggested a phenomenological model including four parameters: (1) mean radius, (2) its standard deviation (or the square root of the zeroth spectral moment), (3) its gradient's standard deviation (or the square root of the second spectral moment), (4) its Hurst exponent and (5) its Nayak parameter.
The model is applicable in a relatively large interval of parameters and properly describes the change in flux with these geometrical parameters.
It is worth noting that the model shows an interplay between the second spectral moment and a specific combination of the Hurst exponent and the Nayak parameter.
The conductivity increases with the former and decreases with the latter.
In the generative model employed in this study, under increasing magnification, 
the second spectral moment and the Nayak parameter increase in such a way that the flux is always a monotonically increasing function (by construction). 
Nonetheless, it is conceivable to design shapes where these two parameters are independently controlled. 
Consequently, the \textit{second question} arises: could an increase in the Nayak parameter actually lead to a reduction in flux in practical scenarios? 
An affirmative response would intriguingly imply the existence of an optimal Nayak parameter (linked to an ideal shape) that maximizes conductivity. 
However, such a scenario seems rather unlikely.

Similar to our analysis of simple multi-petal shapes, the phenomenological model enabled us to determine the ultimate fractal limit for the conductivity of self-affine shapes as the magnification \( \zeta \) approaches infinity. 
This limit depends solely on the mean radius, standard deviation, and only weakly on the Hurst exponent.
However, as in our earlier findings, this identified limit should be regarded as a preliminary estimate. 
The mathematical \textit{question of the conductivity of self-affine shapes in the fractal limit} remains open for further exploration. 
From a physical perspective, similarly to observations with flower-shaped spots, 
the diffusive nature of the boundary could provide a more practical approach to determining this limit.

\subsection{Contact spots between rough surfaces}

\begin{figure}[ht!]
	\centering
	\includegraphics[width=1\linewidth]{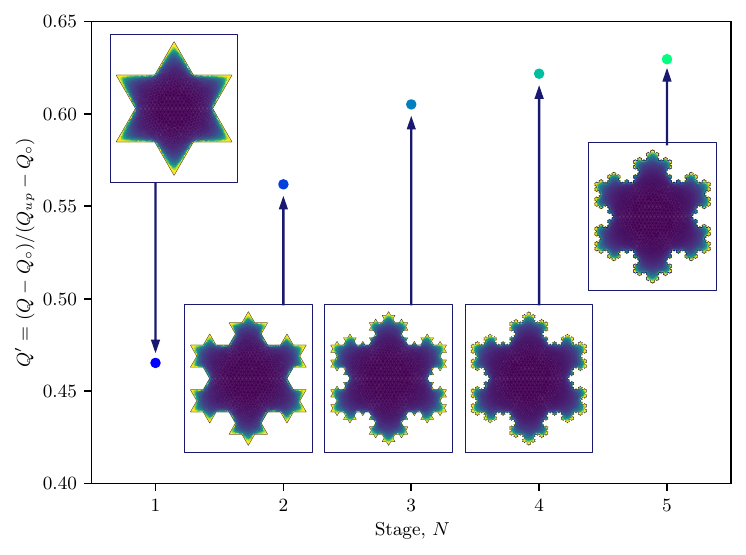}
	\caption{ \label{fig:Koch} Normalized flux at Koch snowflake spots at several first fractal iterations and the evolution of the total flux.}
\end{figure}

Concerning the conductivity of contact spots formed between randomly rough surfaces in contact, we can highlight several pertinent findings.
The non-simple connectedness of these spots, characterized by non-contact areas surrounded by contact ones, does not appear to significantly affect overall conductivity.
However, the complexity of their shapes undoubtedly influences this conductivity. 
Drawing from our analyses of relatively simpler cases, a set of parameters proves effective for estimating conductivity using the developed phenomenological models 
(Eqs.~\eqref{eq:q_selfaffine_model} and ~\eqref{eq:Q_fit_int}). 
These parameters include (1) average radius, (2) standard deviation, (3) second spectral moment, (4) Hurst exponent, and (5) Nayak's parameter of the outer contour.
But this model should be applied to realistic contact spots with caution.
In most cases, such spots cannot be parametrized through a function in polar coordinates $r(\theta)$ and, in general, polar coordinates do not make much sense for complex spots (see Fig.~\ref{fig:0}).
Instead, a more general parametrization using convective coordinates defined along the outer boundary is needed.
In this study, by limiting ourselves to relatively simple geometrical models, we left more realistic contact shapes for future research.
As a preliminary simulation for this research, we present a conductivity analysis of Koch snowflakes~\cite{koch1904courbe}, 
with conductivity results for several initial iterations shown in Fig.~\ref{fig:Koch}. 
Qualitatively, these results align with the observed conductivity saturation at the fractal limit, a natural outcome for the physical problem of conduction.

\section*{Acknowledgement}

This work is a part of PhD project of PB which was funded by MINES Paris - PSL as "Th\`ese-Open".
The authors are grateful to Samuel Forest and Cristian Ovalle-Rodas for thoughtful discussions and for their useful remarks during the whole PhD project of PB.
The authors also would like to thank St\'ephanie Chaillat for her guidance with hierarchical matrices and Basile Marchand for his help with programming issues.
In this work, 
a Large Language Model assisted the authors with polishing the text and, to a little extent, with the code improvement.

\appendix

\section{\label{app:apc}Geometrical characteristics of flower-, star- and gear-like shapes}

\begin{figure}[ht!]
	\centering
	\includegraphics[width=1\linewidth]{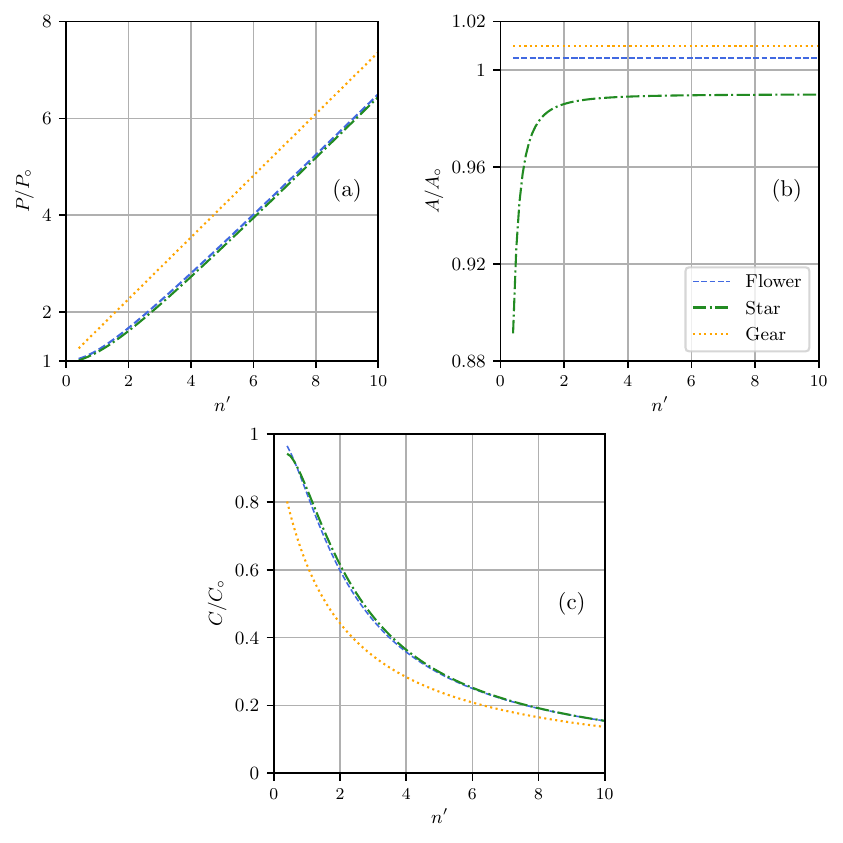}
	\caption{\label{fig:apc} Perimeter, area and compactness for the different shapes of contact spot}
\end{figure}

 Geometrical characteristics (perimeter, area and compactness) are summarized in Table~\ref{tab:geom} and Fig.~\ref{fig:apc} for {flower-,} {star-} and gear-like spots.
 We remind that the mean radius is denoted by $r_0$, and half-length of petals (stars or teeth) is equal to $r_0\xi$.

 \begin{table}
    \begin{tabular}{l|ccc}
         Characteristics & Flower & Star & Gear \\
     \hline\\[3pt]
     Area $A$ & 
     $\pi r_0^2 \left( 1 + \frac{\xi^2}{2} \right)$ & 
     $n r_0^2 ( 1 - \xi^2 ) \sin( \pi / n)$  & 
     $\pi r_0^2 (1 + \xi^2 )$\\[7pt]
     Perim. $P$ & 
     $r_0 E(in')$ & 
     $2 \sqrt{2} n r_0 \sqrt{ 1 + \xi^2 - ( 1 - \xi^2 ) \cos( \pi / n ) }$ & 
     $2 \pi r_0 ( 1 + 2 n' / \pi )$\\[7pt]
     Comp. $C = \frac{\sqrt{A}}{P}$ & 
     $\displaystyle\sqrt{\pi}\frac{\sqrt{1+\xi^2/2}}{4E(in')}$ & 
     $\displaystyle\frac{ \sqrt{( 1 - \xi^2 ) \sin( \pi / n)}}{ 2 \sqrt{2n} \sqrt{ 1 + \xi^2 - ( 1 - \xi^2 ) \cos( \pi / n) } }$ & 
     $\displaystyle\frac{\sqrt{ ( 1 + \xi^2 )}}{2 \sqrt{\pi} ( 1 + 2 n' / \pi )}$ \\[3pt]
    \end{tabular}
    \caption{\label{tab:geom}Geometrical characteristics (area, perimeter and compactness) for multi-petal shapes: flower, star and gear; $E(in') = E(in\xi) = \int\limits_{0}^{\pi/2} \sqrt{1 + \left( n \xi \sin( \theta ) \right) ^2} d \theta$ is the complete elliptic integral of the second kind and $i$ is the imaginary unit.}
 \end{table}

\section{\label{app:geom_selfaffine}Geometrical characteristics of self-affine spots}

\begin{figure}[htb!]
    \centering
  \includegraphics[height=6.2cm]{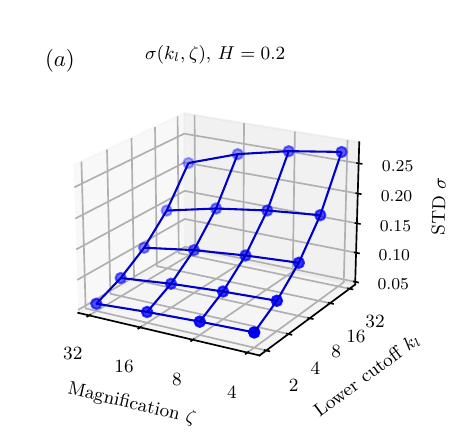}
  \includegraphics[height=6.2cm]{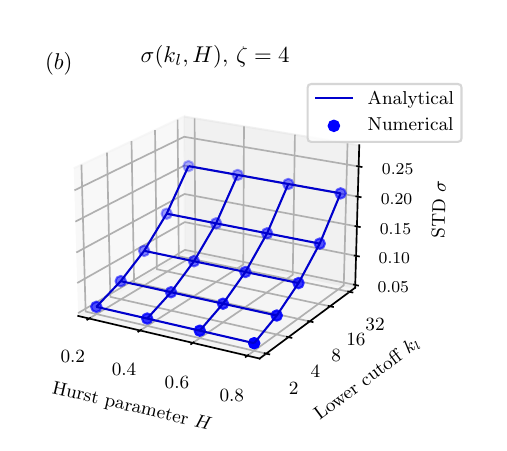}
    \caption{\label{fig:sig_surface}Results of standard deviation by spectral and sample analysis, with $k_l = \{ 2, 4, 8, 16, 32, 64, 128 \}$ in both figures: (a) $\zeta = \{4, 8, 16, 32 \}$ and $H=0.2$; (b) $H = \{0.2, 0.4, 0.6, 0.8 \}$ and $ \zeta = 4$.}
\end{figure}

In Fig.~\ref{fig:sig_surface}, the analytical form for the standard deviation of self-affine spots $\sigma_{s}$ Eq.\eqref{eq:mu_sig_log} is compared with the one evaluated over $1\,000$ generated spots for each combination of generative parameters: $\xi=0.05$, $k_l \in \{ 2, 4, 8, 16, 32, 64, 126 \}$, $\zeta \in \{ 4, 8, 16, 32 \}$ and $H \in [0.2, 0.8]$.
These results are quite sensitive to $k_l$, but the maximal relative error is $0.05 \%$ for $k_l=32$ and $\zeta = 8$.

\begin{figure}[htb!]
    \centering 
    \includegraphics[height=6.2cm]{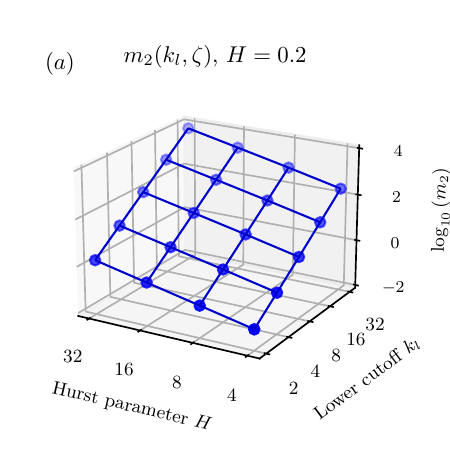}
    \includegraphics[height=6.2cm]{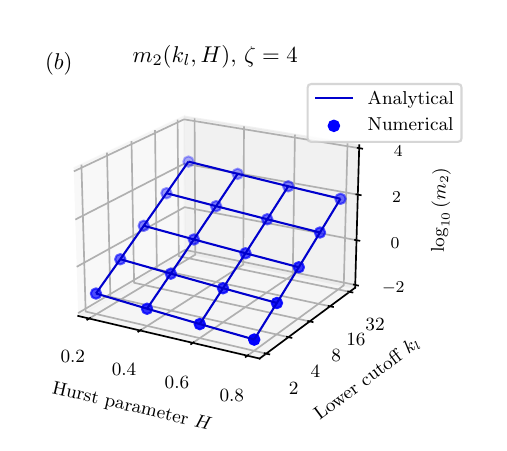}
    \caption{\label{fig:m2_surface}Results of mean square gradient by spectral and sample analysis, with $k_l = \{ 2, 4, 8, 16, 32, 64, 128 \}$ in both figures: (a) $\zeta = \{4, 8, 16, 32 \}$ and $H=0.2$; (b) $H = \{0.2, 0.4, 0.6, 0.8 \}$ and $ \zeta = 4$. }
\end{figure}

\begin{figure}[htb!]
    \centering
    \includegraphics[height=6.2cm]{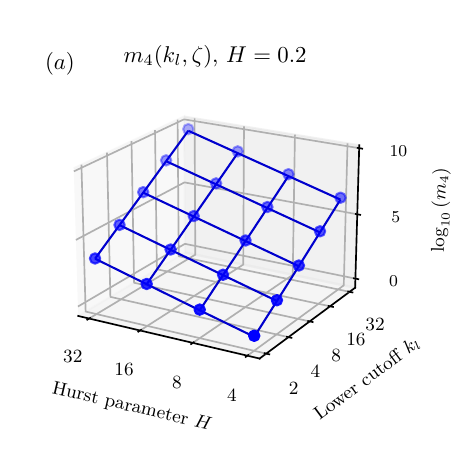}
    \includegraphics[height=6.2cm]{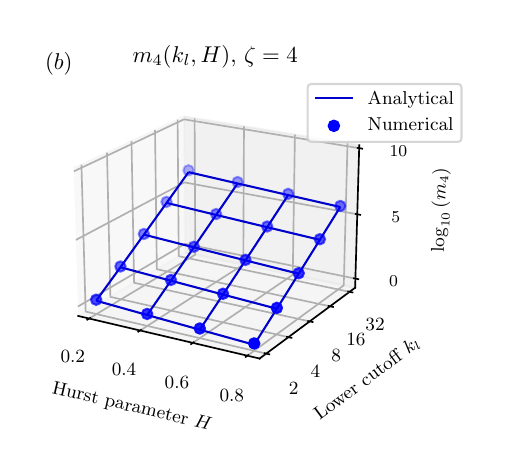}
    \caption{\label{fig:m4_surface}Results of mean square Laplacian, with $k_l = \{ 2, 4, 8, 16, 32, 64, 128 \}$ in both figures: (a) $\zeta = \{4, 8, 16, 32 \}$ and $H=0.2$; (b) $H = \{0.2, 0.4, 0.6, 0.8 \}$ and $ \zeta = 4$.}
\end{figure}

The second $m_2$ and the forth $m_4$ moments have been computed for the same set of generative parameters over the same $1\,000$ spots.
These moments could be computed in three different ways.
First, the discretized contour geometry could be used to evaluate these moments $m_p^D$ using Eq.~\eqref{eq:m0_m2_m4}. 
The discretization consists of splitting the contour in $N = \max\{10\,000, 100k_s\}$ straight segments with $d\theta = 2\pi/N$ and evaluation gradient and Laplacian as 
\begin{equation}
 \nabla r_i = \frac{2(r_{i+1} - r_{i})}{(r_{i+1} + r_i) d \theta}, \quad \Delta r_i =   \frac{4(r_{i+1} - 2r_{i} + r_{i-1})}{(r_{i+1} + 2r_i + r_{i-1})^2 d \theta^2},
\end{equation}
where $r_i = r(i d\theta),\; i = 1,N$. This method was used throughout the paper.
Second, the moments could be approximated by discrete sums of all mode contributions as in Eq.~\eqref{eq:m0_m2_m4_analytical}:
\begin{equation}
	m_p^{S} = \frac{(r_0\xi)^2}{2} \sum_{k_l}^{k_s} k^p {\left( \frac{k}{k_l} \right) }^{-(1+2H)}.
	\label{eq:mps}
\end{equation}
This method is however valid only for relatively small values of $\xi$ because it ignores the exponential transformation~\eqref{eq:rregink}.
Third, for sufficiently large values of $k_l$, these discrete sums could be turned into integrals with wavenumber $k$ becoming a continuum variable of integration:    
\begin{equation}
	m_p^{C} = \frac{1}{2} \int_{k_l}^{\zeta k_l} \left(r(k)\right)^2 k^p dk,
	\label{eq:mpc}
\end{equation}
where $r(k) =  \xi r_0 (k/k_l)^{-H-0.5}$.
Analytical formulas derive from the development for the moments $m_0^{C}$, $m_2^{C}$ and $m_4^{C}$, as follows.
\begin{equation}
	\label{eq:mpc_set}
	m_0^{C} = - \frac{(r_0\xi)^2 k_l}{4 H} \left( \zeta^{-2H} -1 \right), \quad 
	m_2^{C} = - \frac{(r_0\xi)^2 k_l^3}{2(2-2H)} \left( \zeta^{2-2H} -1 \right), \quad 
	m_4^{C} = - \frac{(r_0\xi)^2 k_l^5}{2(4-2H)} \left( \zeta^{4-2H} -1 \right).
\end{equation}
For high values of $k_l$ the discrete spectrum is closer to a continuous one, and thus the spectral moments can be deduced from Eq.~\eqref{eq:m0_m2_m4_analytical} as detailed by Nayak~\cite{nayak1971random}.
These analytical values are compared with the numerically evaluated ones in Fig.\ref{fig:m2_surface} and Fig.\ref{fig:m4_surface}.
The maximum deviation is of only $0.7$ \% for the $m_2$ for parameters $k_l = 2, \zeta = 4$ and $H = 0.2$.
However, for $m_4$ an average discrepancy of $10$ \% is observed and could raise to as much as $27$ \% in certain instances.
Nevertheless, in all results presented in the paper only actual values of the moments and of their combinations were used.

\begin{figure}[ht!]
    \centering
    \includegraphics[width=1\textwidth]{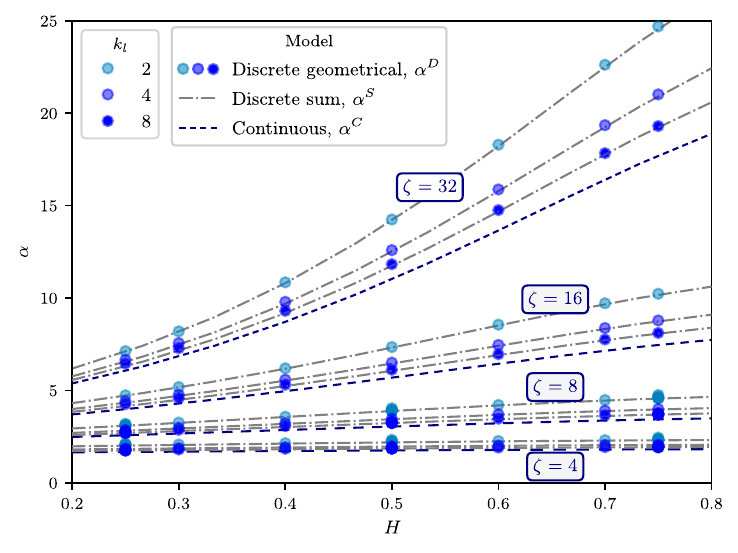}
    \caption{ \label{fig:alpha_model} Comparison of different models which could be used to evaluate spectral moments and the Nayak parameter: (1) discretized geometrical evaluation, (2) discrete sum for the generative function, (3) continuous version of this discrete sum.}
\end{figure}

The Nayak's parameter is determined using the moments $m_0$, $m_2$ and $m_4$, $\alpha = m_0m_4/m_2^2$.
The three models mentioned above could be used to compute the Nayak parameter as $\alpha^D$, $\alpha^S$ and $\alpha^C$, respectively.
The average values computed  over a set of $1\,000$ spots are compared in Fig~\ref{fig:alpha_model}.
This analysis demonstrates that a continuum model could be successfully used in practical applications.
In the limit of high magnification $\zeta$, the second moment and the Nayak parameter scale as $m_2^C \sim \zeta^{2-2H}$ and $\alpha^C \sim \zeta^{2H}$.

\section{\label{app:dQ} Physical consistency of the phenomenological model}

\begin{figure}[ht!]
	\centering
	\includegraphics[width=1\textwidth]{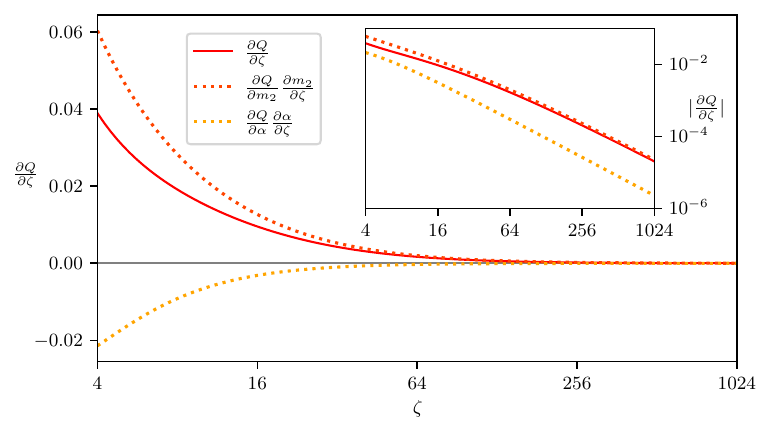}
	\caption{ \label{fig:dQ_dzeta} Derivatives of the flux phenomenological model in function of the magnification $\zeta$ for $k_l=8$, $H = 0.25$, and $\xi = 0.05$ for parameters from Table~\ref{tab:tab_fit}.}
\end{figure}

The phenomenological model of flux Eqs.~\eqref{eq:q_selfaffine_model},\eqref{eq:Q_fit_int} exhibits an increasing behavior with respect of $m_2$, but decreases with $\alpha$. Nevertheless, from general physical considerations, we conjecture that the flux should be a monotonically non-decreasing function of the magnification $\zeta$.
So, we require that the derivative of flux $Q$ with respect to magnification $\zeta$ remains non-negative:
\begin{equation} 
	\frac{ \partial Q}{ \partial \zeta } = 
	\frac{ \partial Q }{ \partial m_2 } \frac{ \partial m_2 }{ \partial \zeta } +
	\frac{ \partial Q }{ \partial \alpha } \frac{ \partial \alpha }{ \partial \zeta } \ge 0
\end{equation}
The terms $ \partial Q / \partial m_2$ and $ \partial Q / \partial \alpha $ could be redily derived from Eq.~\eqref{eq:Q_fit_int}.
\begin{subequations}
	\begin{align}
		\frac{ \partial Q }{ \partial m_2 } &= a \frac{ b }{ 2 \sqrt{m_2} ( b \sqrt{m_2} + 1 )^2 } (1 + c H)\left( 1 + \frac{d}{e \alpha^f +1 } \right)
		\label{eq:dQ_dm2}
	\end{align}
	\begin{align}
		\frac{ \partial Q }{ \partial \alpha } &= a \left\lbrace 1 - \frac{ b }{ b \sqrt{m_2} + 1 } \right\rbrace (1 + c H) \frac{-d e f \alpha^{f-1} }{( e \alpha^f +1)^2 }
		\label{eq:dQ_dalpha}
	\end{align}
\end{subequations}
The derivatives of $m_2$ and $\alpha$ with respect to $\zeta$ could be found from Eq.(\ref{eq:mpc_set}), resulting in the following asymptotic forms:
\begin{subequations}
	\begin{align}
		\frac{ \partial Q }{ \partial m_2 } \frac{ \partial m_2}{ \partial \zeta } &\sim \frac{ 1 }{ \sqrt{m_2} { \left( \sqrt{m_2} + 1 \right) }^2 } \frac{ \partial m_2 }{ \partial \zeta } \sim \zeta^{-2+H}
		\label{eq:sim_dQ_dm2}
	\end{align}
	\begin{align}
		\frac{ \partial Q }{ \partial \alpha } \frac{ \partial \alpha }{ \partial \zeta } &\sim \frac{ \alpha^{f-1} }{ { \left( \alpha^f + 1 \right) }^2 } \frac{ \partial \alpha }{\partial \zeta} \sim \zeta^{-2fH-1}
		\label{eq:sim_dQ_dalpha}
	\end{align}
\end{subequations}
These expressions enable us to define a constraint criterion to ensure the derivative of the flux law with respect to $\zeta$ remains non-negative for all $\zeta$.
The exponent of $\zeta$ in Eq.~\eqref{eq:sim_dQ_dm2} must be lower than the one in Eq.(\ref{eq:sim_dQ_dm2}), resulting in the following inequality that the exponent $f$ should satisfy:
\begin{equation}
	f \geq \frac{1-H}{2H}
	\label{eq:f_bound}
\end{equation}
The problem with this constraint is that it results in too high values of $f$ for small $H$ and, ultimately, it diverges for $H\to 0$.
In the current study we set the minimal value of the Hurst exponent to $H = 0.25$ thus resulting in $f\ge1.5$.
The results for derivatives using the continuous expressions for $m_2$ and $\alpha$ are presented Fig.~\ref{fig:dQ_dzeta} for $H = 0.25$, $k_l=8$, and $\xi = 0.05$, and the fitting parameters shown in Table~\ref{tab:tab_fit}.
The two derivative terms are distinguished: one positive, as given by Eq.~\eqref{eq:dQ_dm2}, and the other negative, as given by Eq.~\eqref{eq:dQ_dm2}.
The full derivative remains positive, however, thus keeping the required assumptions true, even for value of $\zeta$ significantly far from the initial set of parameter.
The absolute values for these derivatives are also depicted in inset in log-log scale, showing similar power-laws of the two competing derivatives (in dots) for high values of $\zeta$.

\end{document}